\documentclass[final,11pt]{elsarticle}%
\usepackage{lineno,hyperref}
\usepackage{graphicx}
\usepackage{subfig}
\usepackage{amsmath,bm}
\usepackage{amsfonts}
\usepackage{amssymb}
\usepackage{amsbsy}
\usepackage{color}
\usepackage{multirow}
\usepackage{framed}
\usepackage{textcomp}
\usepackage{eurosym}
\usepackage{leftidx}
\usepackage{mathtools}
\usepackage{empheq}
\usepackage{booktabs}
\usepackage[top=30mm,left=30mm,body={155mm,210mm}]{geometry}
\usepackage{algorithm,algorithmic}
\usepackage{tikz,pgfplots, pgfplotstable}
\usepackage{mdframed}
\usepackage{newfloat}
\usepackage{amsmath}%
\providecommand{\U}[1]{\protect\rule{.1in}{.1in}}
\pgfplotsset{compat=1.3}
\makeatletter
\renewcommand{\ALG@name}{Box}
\makeatother
\DeclareFloatingEnvironment[
fileext=lob,
listname={List of Boxes},
name=Box,
placement=htp,
]{BOX}

\journal{Computational Mechanics}
\begin{document}

%Define pgfplot style
\pgfplotsset{
legend style = {font=\scriptsize},
ticklabel style = {font=\scriptsize},
label style = {font=\footnotesize},
width = 0.55\textwidth,
height = 0.45\textwidth,
}

\begin{frontmatter}
\title{Thermodynamically consistent nonlinear viscoplastic formulation with well-conditioned recovery of the inviscid solution: Theory and implicit integration algorithm with exact solution for the linear case}
\author[UPM]{K. Nguyen}
\ead{khanhnguyen.gia@upm.es}
\address[UPM]{Escuela T\'ecnica Superior de Ingenier\'ia Aeron\'autica y del Espacio, Universidad Polit\'ecnica de Madrid, Pza. Cardenal Cisneros, 28040, Madrid}
\author[UPM]{V\'ictor J.  Amores}
\ead{victorjesus.amores@upm.es}
\author[UPM]{Francisco J. Mont\'ans }
\ead{fco.montans@upm.es}
\begin{abstract}
In this work, a consistent viscoplasticity formulation is derived from thermodynamical principles and employing the concept of continuum elastic corrector rate. The proposed model is developed based on the principle of maximum viscoplastic dissipation for determining the flow direction. The model uses both the equivalent viscoplastic strain and its rate as state variables.
Power balance and energy balance give, respectively, separate evolution equations for the equivalent viscoplastic strain rate and the viscoplastic strain, the former written  in terms of inviscid rates. Several key points distinguish our formulation from other proposals. First, the viscoplastic strain rate  (instead of a yield function) consistently  distinguishes conservative from dissipative behaviours during reverse loading; and the discrete implicit integration algorithm is an immediate implementation of the continuum theory based on the mentioned principles. Second, the inviscid solution is recovered in a well-conditioned manner by simply setting the viscosity to zero. Indeed, inviscid plasticity, viscoelasticity and viscoplasticity are particular cases of our formulation and integration algorithm, and are recovered just by setting the corresponding parameters to zero (viscosity or yield stress). Third, the linear viscoplasticity solution is obtained in an exact manner for proportional loading cases, independently of the time step employed. Four, general nonlinear models (Perzyna, Norton, etc) may be immediately incorporated as particular cases both in the theory and the computational implementation.
\end{abstract}
\end{frontmatter}

%\linenumbers

\section{Introduction}

The elastoplastic behavior of materials have a time-dependent component,
meaning that the speed at which plastic dissipation takes place affects the
observed behavior. This time-dependent effect is usually modelled through a
viscoplastic constitutive relation. In many cases, when the rate of loading is
very small and the time-dependent effect can be neglected, the
rate-independent elastoplasticity models can provide a good approximation to
the experimental results \cite{KojicBathe,Lubliner}. However, in the cases
when such conditions are not met, the rate-dependency is important, and must
be taken into account in the constitutive model to obtain accurate
predictions. In a general purpose model, the importance of such effects cannot
be determined apriori, so a smooth transition in the simulations from
rate-independent to rate-dependent plasticity is desired. Viscoplasticity is
the common type of model incorporating strain-rate dependent plastic flow.
Furthermore, it is desirable to also incorporate viscoelasticity in the same framework.

Many constitutive viscoplastic models have been presented, including their
dedicated computational treatments. In general, the viscoplasticity models can
be classified into two families. One is the so-called \textit{overstress}
models; the other family comprises the so-called \textit{consistency} models.
The first family is based on the ideas proposed by Perzyna \cite{Perzyna1966},
in which the current stress state can be outside the yield surface and the
yield function may be greater than zero (hence, the \textit{overstress} name).
In these cases, the Kuhn-Tucker conditions typical of plasticity are not
applicable. The rate of an equivalent (visco-)plastic strain $\dot{\gamma}$ is
obtained from a direct evolution equation in terms of the overstress and the
viscosity $\eta$; this rate is incrementally integrated to obtain the
equivalent (visco-)plastic strain $\gamma$. The Perzyna model
\cite{Perzyna1966} and the Duvaut-Lions model \cite{Duvaut1972}, among others,
are the most popular formulations in this first family. Both models are not
only widely used in small strain problems
\cite{Zienkiewicz1974,Hughes1978,Cormeau1975,Simo1988,Chaboche1989,Peric1993,Ristinmaa1998,Runesson1999,Caggiano2018}%
, but have also been extended to finite strain problems
\cite{Ibrahimbegovic2000,Nedjar2002a,Shutov2008,Kowalczyk-Gajewska2019} and
are common also in crystal plasticity, often tailored and referred to as
power-laws \cite{Wang,Miehe}. Nonetheless, despite the improvements and
advances in their computational treatments
\cite{Zienkiewicz1974,Hughes1978,Cormeau1975,Simo1988,Chaboche1989,Peric1993,Ristinmaa1998,Caggiano2018}%
, both models still present limitations. The major drawback of the Perzyna
model is that this model has an ill-conditioned inviscid limit \cite{Simo1998}
and because of its inherent structure, it may not naturally converge to the
inviscid solution when the viscosity tends to zero for non-smooth
multi-surface viscoplasticity \cite{Simo1988,Souza-Neto}, a key aspect in
crystal plasticity. The Duvaut-Lions model has the advantage
compared with the Perzyna model in that it can be combined with a non-smooth
yield surface, and the formulation naturally incorporates the inviscid limit
as part of the solution. In this model, the trial and the inviscid solutions
are computed first and then the viscous solution is determined as a relaxation
of the trial state to the inviscid solution, a relaxation which depends on the
characteristic (relaxation) time. However, the advantage is sometimes seen as a handicap respect to Perzyna's model,
because it must be used in conjunction with a separate integration algorithm
for the inviscid elastoplastic rate equations, where the evolution rule is
needed for the yield surface, in case of hardening or softening plasticity
\cite{Simo1988}. But more importantly, in principle the Duvaut-Lions model
does not incorporate general relations of the Perzyna type, being restricted
to linear viscoplasticity, so it is seldom used when the rate-independent
solution is not important and the viscous contribution is expected to be relevant.

The second family of viscoplasticity models has been introduced by Wang et al.
\cite{Wang1997} and then further explored by many authors
\cite{Ristinmaa2000,Carosio2000,Heeres2002,Zaera2006}. This approach includes
the viscoplastic behaviour by incorporating the time-dependency in a so-called
\textit{rate-dependent yield surface}; the purpose being that the Kuhn-Tucker
conditions, typical of rate independent plasticity, remain valid. The
viscoplastic multiplier is determined from a non-homogeneous differential
equation derived from the consistency condition at the rate-dependent yield
surface, so these models are referred to as the \textquotedblleft consistency
models\textquotedblright. The elastic domain in the stress $\boldsymbol{\sigma
}$-space is defined as $\mathbb{E}_{\boldsymbol{\sigma}}=\{\boldsymbol{\sigma
}\in\mathbb{S}~|~f(\boldsymbol{\sigma},\gamma,\dot{\gamma})\leq0\}$, meaning
that in the unloading case, the consistency model always unloads elastically
\cite{Heeres2002} and $f(\boldsymbol{\sigma},\gamma,\dot{\gamma})=0$ is the
viscoplastic yield function. This implies that the rate-dependent yield
surface remains fixed during the unloading phase; in other words, the
viscoplastic multiplier ($\dot{\gamma}$) is not changed during unloading and
is greater than zero (see e.g. Secs. 2.2 and 3.2 of \cite{Heeres2002}). In
essence, this type of models presents the contradiction that at unloading
detected by $f(\boldsymbol{\sigma},\gamma,\dot{\gamma})\leq0$, plastic flow
stops suddenly producing conservative behaviour with frozen $\dot{\gamma}>0$,
values which are inherent to a dissipative process. Hence, these formulations
seem just motivated by numerical difficulties, but result in contradictory
physical conditions.

In this paper, we introduce a novel thermodynamically motivated consistent
viscoplastic formulation which naturally includes a well-conditioned recovery
of the inviscid solution by simply setting the viscosity $\eta=0$. The model
avoids the limitations of the previous models, but incorporates their
advantageous features, including general nonlinear viscosities and hardening.
Furthermore, our proposal is not just a numerical convenience, but it is
motivated in a proper implementation of physical principles. Indeed, our
proposal is postulated from the principle of maximum dissipation in a
straightforward manner, from which a function $f(\boldsymbol{\sigma}%
,\gamma,\dot{\gamma})$ is obtained as a consequence of power conservation (not
from a postulate) to include the rate dependence. Power balance and energy
balance give, respectively, separate evolution equations for the viscoplastic
strain rate and for the viscoplastic strain. This separation allows for the
integration of plasticity, viscoplasticity, and viscoelasticity in a single
computational setting, because plastic strain evolution and its rate are
different variables with their own evolution equations, each one dominating
the particular cases of inviscid plasticity or viscoelasticity. Unlike the
consistency model proposed by Wang et al. \cite{Wang1997}, in our model the
trial viscoplastic multiplier $\dot{\gamma}$ is used consistently to check
whether either dissipation or conservative behavior occurs. As a result,
dissipation can still be generated during the ``unloading'' phase
($f(\boldsymbol{\sigma},\gamma,\dot{\gamma})<0$), until $\dot{\gamma}$
vanishes, even when the trial state lies inside the inviscid yield function. This viscoplastic rate is obtained from an evolution equation in
rate form in terms of inviscid rates. Whereas in the continuum theory we show
that power balance results in energy balance by integration, in the discrete
general theory, both principles facilitate different equations to compute
$\Delta\dot\gamma$ and $\Delta\gamma$. The formulation may accommodate most of
the nonlinear uniaxial viscoplastic models such as Perzyna, Duvaut-Lions and
Norton-type power laws, etc.

An implicit integration algorithm derived immediately from the continuum
theory, based on the novel framework employing continuum elastic rate
correctors, is also proposed including general nonlinear viscoplasticity
\cite{LatMonPlas}. The exact solution, independent of the time increment
employed, is recovered for linear small strain $J_{2}$--viscoplasticity under
proportional loading (as for the case of linear elastoplasticity). We compare
results with some of the well-known viscoplastic models such as the Perzyna,
the Duvaut-Lions and the consistency models. We focus on the ideas behind the
proposal, so we employ in the presentation infinitesimal strains. A large
strains implementation using a framework with logarithmic strains, a
multiplicative decomposition of the deformation gradient and the continuum
elastic corrector rates framework is simple, being the algorithmic difficulty
just related to the kinematic mappings, see e.g.
\cite{SanzLatMon,Sanzetal,ZhangMontans} for this type of formulations, and
\cite{NguyenSanzMontans} for a simple large-strain plane-stress implementation
of this type of approach. Finally, finite element non-homogeneous numerical
examples are presented using our model to demonstrate its numerical
implementation and the computational efficiency of our proposal.

\section{Derivation of the model from thermodynamic principles}

\subsection{Dissipation inequality}

\begin{figure}[pth]
\centering
\includegraphics[width=0.5\textwidth]{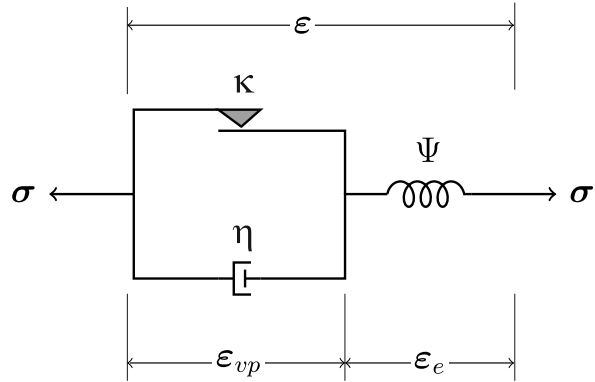}\caption{Rheological
model for viscoplasticity.}%
\label{fig.binghammodel}%
\end{figure}

In this section we establish the basic equations of the consistency
viscoplastic model based on the rheological model shown in Figure
\ref{fig.binghammodel}. This rheological model is well-known as the Bingham
model, which motivates many viscoplastic formulations. Noteworthy, the Bingham
model recovers the Maxwell viscoelasticity rheological model if the yield
stress vanishes, and it recovers the Prandtl plasticity rheological model if
the viscosity vanishes. Then, such cases should be naturally recovered both by
the continuum theory and by the integration algorithm simply setting the
respective constants to zero. Unfortunately, this is not the usual case in the
literature.

The rheological model element considers two strain-like internal variables,
$\boldsymbol{\varepsilon}_{e}$ as the elastic strain governing the
conservative behaviour through $\Psi_{int}(\boldsymbol{\varepsilon}_{e}%
)\equiv\Psi(\boldsymbol{\varepsilon}_{e})$ and $\boldsymbol{\varepsilon}_{vp}%
$, as a viscoplastic strain common to both the friction and the damper element
and, hence, governing the dissipative behaviour. It also considers an external
strain variable $\boldsymbol{\varepsilon}$, a result of the external work. We
focus on conservation principles, so we consider the explicit dependencies
given by $\boldsymbol{\varepsilon}_{e}\left(  \boldsymbol{\varepsilon
},\boldsymbol{\varepsilon}_{vp}\right)  $, which results, by straightforward
use of the chain rule, in%
\begin{equation}
\boldsymbol{\dot{\varepsilon}}_{e}=\frac{\partial\boldsymbol{\varepsilon}_{e}%
}{\partial\boldsymbol{\varepsilon}}:\boldsymbol{\dot{\varepsilon}}%
+\frac{\partial\boldsymbol{\varepsilon}_{e}}{\partial\boldsymbol{\varepsilon
}_{vp}}:\boldsymbol{\dot{\varepsilon}}_{vp}=:~^{tr}\boldsymbol{\dot
{\varepsilon}}_{e}+~^{ct}\boldsymbol{\dot{\varepsilon}}_{e} \label{eq1}%
\end{equation}
where, note, $~^{tr}\boldsymbol{\dot{\varepsilon}}_{e}$ and $~^{ct}%
\boldsymbol{\dot{\varepsilon}}_{e}$ refer, respectively, to trial and
corrector \emph{continuum rates} of the elastic strain, not to algorithmic
ones. The infinitesimal strains in this presentation, based on elastic
corrector rates, facilitate an immediate extension to finite strains based on
the multiplicative decomposition preserving the additive structure; see
\cite{LatMonPlas,SanzLatMon,Sanzetal}. If $\mathcal{P}=\boldsymbol{\sigma
}:\boldsymbol{\dot{\varepsilon}}$ is the external power and $\dot{\Psi}$ is
the change rate of the stored energy, by definition, the dissipation power is%
\begin{align}
\left.  \mathcal{D}^{p}\equiv\mathcal{P}-\dot{\Psi}\right.   &
=\boldsymbol{\sigma}:\boldsymbol{\dot{\varepsilon}}-\frac{d\Psi\left(
\boldsymbol{\varepsilon}_{e}\right)  }{d\boldsymbol{\varepsilon}_{e}%
}:\boldsymbol{\dot{\varepsilon}}_{e}\\
&  =\boldsymbol{\sigma}:\boldsymbol{\dot{\varepsilon}}-\frac{d\Psi\left(
\boldsymbol{\varepsilon}_{e}\right)  }{d\boldsymbol{\varepsilon}_{e}}:\left(
~^{tr}\boldsymbol{\dot{\varepsilon}}_{e}+~^{ct}\boldsymbol{\dot{\varepsilon}%
}_{e}\right)  \geq0 \label{eq3}%
\end{align}
where in Eq. (\ref{eq3}) we used Eq. (\ref{eq1}). Now, following the typical
Coleman arguments \cite{Truesdell}, we analyse the two different cases
(namely, the conservative and dissipative components of the power
$\mathcal{P}=\dot\Psi+ \mathcal{D}^{p}$):

\begin{itemize}
\item \emph{Conservative case}: In the case of absence of dissipation,
$\boldsymbol{\dot{\varepsilon}}_{vp}=\boldsymbol{0}=~-~^{ct}\boldsymbol{\dot
{\varepsilon}}_{e}$, and
\begin{equation}
\mathcal{D}^{p}=\boldsymbol{\sigma}:\boldsymbol{\dot{\varepsilon}}-\frac
{d\Psi\left(  \boldsymbol{\varepsilon}_{e}\right)  }{d\boldsymbol{\varepsilon
}_{e}}:\frac{\partial\boldsymbol{\varepsilon}_{e}}{\partial
\boldsymbol{\varepsilon}}:\boldsymbol{\dot{\varepsilon}}\equiv0
\end{equation}
which must hold for any arbitrary $\boldsymbol{\dot{\varepsilon}}$, so
necessarily---note the abuse of notation in keeping the same symbol for the
functions regardless of their arguments%
\begin{equation}
\boldsymbol{\sigma}=\frac{\partial\Psi(\boldsymbol{\varepsilon}%
,\boldsymbol{\varepsilon}_{vp})}{\partial\boldsymbol{\varepsilon}}\equiv
\frac{d\Psi\left(  \boldsymbol{\varepsilon}_{e}\right)  }%
{d\boldsymbol{\varepsilon}_{e}}:\frac{\partial\boldsymbol{\varepsilon}%
_{e}\left(  \boldsymbol{\varepsilon},\boldsymbol{\varepsilon}_{vp}\right)
}{\partial\boldsymbol{\varepsilon}}=\frac{d\Psi\left(  \boldsymbol{\varepsilon
}_{e}\right)  }{d\boldsymbol{\varepsilon}_{e}}=:\boldsymbol{\sigma}^{|e}
\label{Stressdef}%
\end{equation}
where ${\partial\boldsymbol{\varepsilon}_{e}\left(  \boldsymbol{\varepsilon
},\boldsymbol{\varepsilon}_{vp}\right)  }/{\partial\boldsymbol{\varepsilon}%
}=\mathbb{I}^{s}$, the fourth order fully symmetric identity tensor, is due to
the additive setting that governs infinitesimal strains. At large strains this
identification does not necessarily holds, but the concept of elastic
corrector rate and its additive structure using logarithmic strains do,
maintaining unaltered the additive structure of the infinitesimal theory and
related algorithm at large strains
\cite{LatMonViscoCM,LatMonViscoCAS,LatMonPlas}. Note also that in this
infinitesimal case, $\boldsymbol{\sigma}^{|e}:=d\Psi/d\boldsymbol{\varepsilon
}_{e}$ equals the stress tensor $\boldsymbol{\sigma}$ obtained from external
power balance in $\boldsymbol{\sigma}:\boldsymbol{\dot\varepsilon}$.

\item \emph{Purely dissipative case}: Using Eq. (\ref{Stressdef}), the
external power is frozen, i.e. $\boldsymbol{\dot{\varepsilon}}=\boldsymbol{0}%
$, so we have%
\begin{equation}
\mathcal{D}^{p}=-\boldsymbol{\sigma}:~^{ct}\boldsymbol{\dot{\varepsilon}}_{e}%
\end{equation}
Using the constraint of isochoric flow, the principle of maximum dissipation
implies that \cite{ZhangMontans}%
\begin{equation}
~^{ct}\boldsymbol{\dot{\varepsilon}}_{e}=-c\dot{\gamma}\boldsymbol{\hat{n}}
\label{floweq}%
\end{equation}
where $\boldsymbol{\hat{n}=\sigma}^{d}\boldsymbol{/}\left\Vert
\boldsymbol{\sigma}^{d}\right\Vert $ is the associated constrained flow
direction, $\boldsymbol{\sigma}^{d}$ is the deviatoric stress and $\left\Vert
\boldsymbol{\sigma}^{d}\right\Vert $ is its norm, and $c\dot{\gamma}$ is a
multiplier. The constant $c=\sqrt{3/2}$ is the scalar to account for uniaxial
comparison so $\dot{\gamma}$ takes the convenient uniaxial equivalence
meaning; i.e. during a uniaxial test in the $x$--direction%
\begin{equation}
\left(  ~^{ct}\boldsymbol{\dot{\varepsilon}}_{e}\right)  _{x}=-\sqrt{\tfrac
{2}{3}}c\dot{\gamma}\text{ \ so we take \ }c=\sqrt{\tfrac{3}{2}}\text{ \ to
get }\left(  ~^{ct}\boldsymbol{\dot{\varepsilon}}_{e}\right)  _{x}%
=-\dot{\gamma}%
\end{equation}
For the classical infinitesimal case with isochoric flow, denoting the
volumetric strain by $\varepsilon_{v}=tr(\boldsymbol{\varepsilon
})=tr(\boldsymbol{\varepsilon}_{e})$ and the deviatoric elastic one by
$\boldsymbol{\varepsilon}_{e}^{d}=\boldsymbol{\varepsilon}_{e}-\tfrac{1}%
{3}\varepsilon_{v}\boldsymbol{I},$ we consider the stored energy function%
\begin{equation}
\Psi(\boldsymbol{\varepsilon}_{e})=\tfrac{1}{2}2\mu\boldsymbol{\varepsilon
}_{e}^{d}:\boldsymbol{\varepsilon}_{e}^{d}+\tfrac{1}{2}K\varepsilon_{v}^{2}%
\end{equation}
where $\mu$ is the shear modulus and $K$ is the bulk modulus. Using
$d\boldsymbol{\varepsilon}_{e}^{d}/d\boldsymbol{\varepsilon}_{e}%
=\mathbb{P}^{d}$, the deviatoric projector, and $d\varepsilon_{e}%
^{v}/d\boldsymbol{\varepsilon}_{e}=\boldsymbol{I}$, the identity tensor, the
resulting trial stress rate is%

\begin{equation}
~^{tr}\boldsymbol{\dot{\sigma}}\equiv~^{tr}\boldsymbol{\dot{\sigma}}%
^{|e}:=\frac{d\boldsymbol{\sigma}^{|e}}{d\boldsymbol{\varepsilon}_{e}}%
:~^{tr}\boldsymbol{\dot{\varepsilon}}_{e}=\frac{d^{2}\Psi
(\boldsymbol{\varepsilon}_{e})}{d\boldsymbol{\varepsilon}_{e}\otimes
d\boldsymbol{\varepsilon}_{e}}:~^{tr}\boldsymbol{\dot{\varepsilon}}%
_{e}=\underset{%
\begin{array}
[c]{c}%
^{tr}\boldsymbol{\dot{\sigma}}^{d}%
\end{array}
}{\underbrace{2\mu~^{tr}\boldsymbol{\dot{\varepsilon}}_{e}^{d}}}%
+K\dot{\varepsilon}_{v}\boldsymbol{I}%
\end{equation}

and by Eq. (\ref{floweq}), the corrector stress rate is%
\begin{equation}
^{ct}\boldsymbol{\dot{\sigma}}\equiv\,^{ct}\boldsymbol{\dot{\sigma}}%
^{|e}:=\mathbb{C}_{e}:~^{ct}\boldsymbol{\dot
{\varepsilon}}_{e}=2\mu~^{ct}\boldsymbol{\dot{\varepsilon}}_{e}=-2\mu
c\dot{\gamma}\boldsymbol{\hat{n}}%
\end{equation}
where $\mathbb{C}_{e}:=d^{2}\Psi(\boldsymbol{\varepsilon}_{e}%
)/d\boldsymbol{\varepsilon}_{e}\otimes d\boldsymbol{\varepsilon}_{e}$ is the elastic tangent. Because of the deviatoric nature of $\boldsymbol{\hat{n}}$ we have
$\boldsymbol{\sigma}:\boldsymbol{\hat{n}}=\boldsymbol{\sigma}^{d}%
:\boldsymbol{\hat{n}}$. Note that despite that we include herein the familiar
rate forms for the infinitesimal case, the stresses are hyperelastic, i.e.
\begin{equation}
\boldsymbol{\sigma}\equiv\boldsymbol{\sigma}^{|e}(\boldsymbol{\varepsilon}%
_{e}):=\frac{d\Psi(\boldsymbol{\varepsilon}_{e})}{d\boldsymbol{\varepsilon
}_{e}} \label{eqhyper}%
\end{equation}
so stress rate forms bellow are included just to facilitate the reader
comparisons with other infinitesimal formulations. For the finite case, or for
infinitesimal bi-modulus materials \cite{Bimodulus}, direct hyperelastic
relations are more convenient.
\end{itemize}

\subsection{Thermodynamic consistency}

Let us consider the aforementioned Bingham-Maxwell-Prandtl model, where a spring element, representing a stored energy, is in series
with two dissipative elements in parallel (one friction and one damper). In
the absence of external power (which requires $\boldsymbol{\dot{\varepsilon
}=0}$), we must have the following relation from thermodynamic consistency
(i.e. equivalence of the dissipation, or that the dissipated power equals the
decrease rate of the stored energy for the case of frozen external power)%
\begin{equation}
\mathcal{D}^{p}\equiv-\boldsymbol{\sigma}:~^{ct}\boldsymbol{\dot{\varepsilon}%
}_{e}=\kappa\left(  \gamma_{p}\right)  \dot{\gamma}_{p}+g\left(  \dot{\gamma
}_{v}\right)  \dot{\gamma}_{v}\geq0 \label{diseq}%
\end{equation}
where $\gamma_{p}$ is the uniaxial-equivalent plastic strain (the cumulative
sliding in the friction element) and $\dot{\gamma}_{v}$ is the velocity of
displacement in the damper. The functions $\kappa(\gamma_{p})$ and
$g(\dot{\gamma}_{v})$ are the, possibly nonlinear, scalar uniaxial-equivalent
functions representing the energy-conjugate stress-like internal variables in
the friction and the damper elements, respectively. Furthermore, if both
elements are in parallel, it is obvious that the kinematics imply that%
\begin{equation}
\dot{\gamma}_{p}=\dot{\gamma}_{v}\equiv\dot{\gamma}=\frac{d\gamma}{dt}%
\end{equation}
Note that another implication of the description given by the rheological
model is that the dissipation can be decoupled in an additive manner as
described in the previous equations, separating the dependence on $\gamma$
from that on $\dot{\gamma}$. ~With the above definitions and assumptions
motivated from the rehological model, the equal sign identifying both versions
of the dissipation in Eq. (\ref{diseq}), states that
\begin{equation}
-\boldsymbol{\sigma}:~^{ct}\boldsymbol{\dot{\varepsilon}}_{e}-\kappa\left(
\gamma\right)  \dot{\gamma}-g\left(  \dot{\gamma}\right)  \dot{\gamma}=0
\end{equation}
Then, using Eq. (\ref{floweq}), the following two conditions must hold, the
first one implying the first principle of thermodynamics (conservation of
power by the identity in Eq. (\ref{diseq})) and the second one implying the
non-negativity of dissipation from the second principle (the \textquotedblleft%
$\geq$\textquotedblright\ sign in Eq. (\ref{diseq}))%
\begin{equation}
\left\{
\begin{array}
[c]{l}%
f\dot{\gamma}:=[f_{p}(\boldsymbol{\varepsilon}_{e},\gamma)-g(\dot{\gamma
})]\dot{\gamma}:=\left[  c~\boldsymbol{\sigma}:\hat{\boldsymbol{n}}%
-\kappa\left(  \gamma\right)  -g\left(  \dot{\gamma}\right)  \right]
\dot{\gamma}=0\text{ \ \ (first principle)}\\
\;\\
\mathcal{D}^{p}:=\left[  \kappa\left(  \gamma\right)  +g\left(  \dot{\gamma
}\right)  \right]  \dot{\gamma}\geq0\text{ \ (second principle)}%
\end{array}
\right.  \label{conditions}%
\end{equation}
Note that from Eq. (\ref{eqhyper}) we can write the dependencies either using
the elastic strains as in $f_{p}(\boldsymbol{\varepsilon}_{e},\gamma)$ or
using the stress as in $f_{p}(\boldsymbol{\sigma},\gamma)$; recall that to
avoid proliferation of symbols, we use the same symbols for functions with a
same physical meaning, regardless of the arguments (if convenient, we will write
the relevant ones in the discussion, explicitly).

We usually require that the dissipation in both dissipative elements must be
positive by themselves, i.e. $\kappa\left(  \gamma\right)  \dot{\gamma}\geq0$
and $g\left(  \dot{\gamma}\right)  \dot{\gamma}\geq0$, which is guaranteed if
$\kappa\left(  \gamma\right)  \geq0$ and $g\left(  \dot{\gamma}\right)  \geq0$
and $\dot{\gamma}\geq0$. In fact, $\dot{\gamma}\geq0$ is usually considered a
requirement by definition (i.e. $\gamma$ is a monotonically increasing
variable). Then, from the first condition in Eq. (\ref{conditions}), we have

\begin{itemize}
\item if $\dot{\gamma}>0$, which corresponds to a dissipative case, the first
principle implies%
\begin{equation}
f\left(  \boldsymbol{\varepsilon}_{e},\gamma,\dot{\gamma}\right)  \equiv
f_{p}(\boldsymbol{\varepsilon}_{e},\gamma)-g(\dot{\gamma})\equiv
c~\boldsymbol{\sigma}(\boldsymbol{\varepsilon}_{e}):\boldsymbol{\hat{n}%
}(\boldsymbol{\varepsilon}_{e})-\kappa\left(  \gamma\right)  -g\left(
\dot{\gamma}\right)  =0. \label{fdefinition}%
\end{equation}

\item if $\dot{\gamma}=0$, which corresponds to a conservative case, we may
have $f=0$, $f>0$ or $f<0$. Now, we analyze the case that $f>0$, from the fact
that no dissipation is taken place and the viscoplastic strain is frozen
($\dot{\gamma}_{p}=\dot{\gamma}_{v}=0$). We assume that $g(0)=0$; no stress in
the dashpot for $\dot{\gamma}_{v}=0$. Then, the case $f>0$ requires
\begin{equation}
f\left(  \boldsymbol{\varepsilon}_{e},\gamma,\dot{\gamma}\right)  \equiv
f_{p}(\boldsymbol{\varepsilon}_{e},\gamma):=c~\boldsymbol{\sigma
}(\boldsymbol{\varepsilon}_{e}):{\hat{\boldsymbol{n}}}(\boldsymbol{\varepsilon
}_{e})-\kappa(\gamma)>0 \label{fpeq}%
\end{equation}
However, by the definition in the rheological model, $\kappa(\gamma)$ is the
yield stress and by definition of the $\boldsymbol{\hat{n}}$ symbol
$\boldsymbol{\sigma}:{\hat{\boldsymbol{n}}}>0$, so $f_{p}>0$ implies
$c~||\boldsymbol{\sigma}^{d}||>\kappa(\gamma)$. In turn this implies by
equilibrium in the friction element an increment in the plastic strain,
$\dot{\gamma}_{p}>0$, which would be in contradiction with our original
assumption for this case. Consequently by the definition of $\kappa(\gamma)$,
the condition $\dot{\gamma}=0$ requires $f=f_{p}\leq0$ and the condition $f>0$ is not
possible. Note that this condition is coincident with that of the inviscid
(purely plastic) case.
\end{itemize}

\begin{figure}[pth]
\centering
\includegraphics[width=0.7\textwidth]{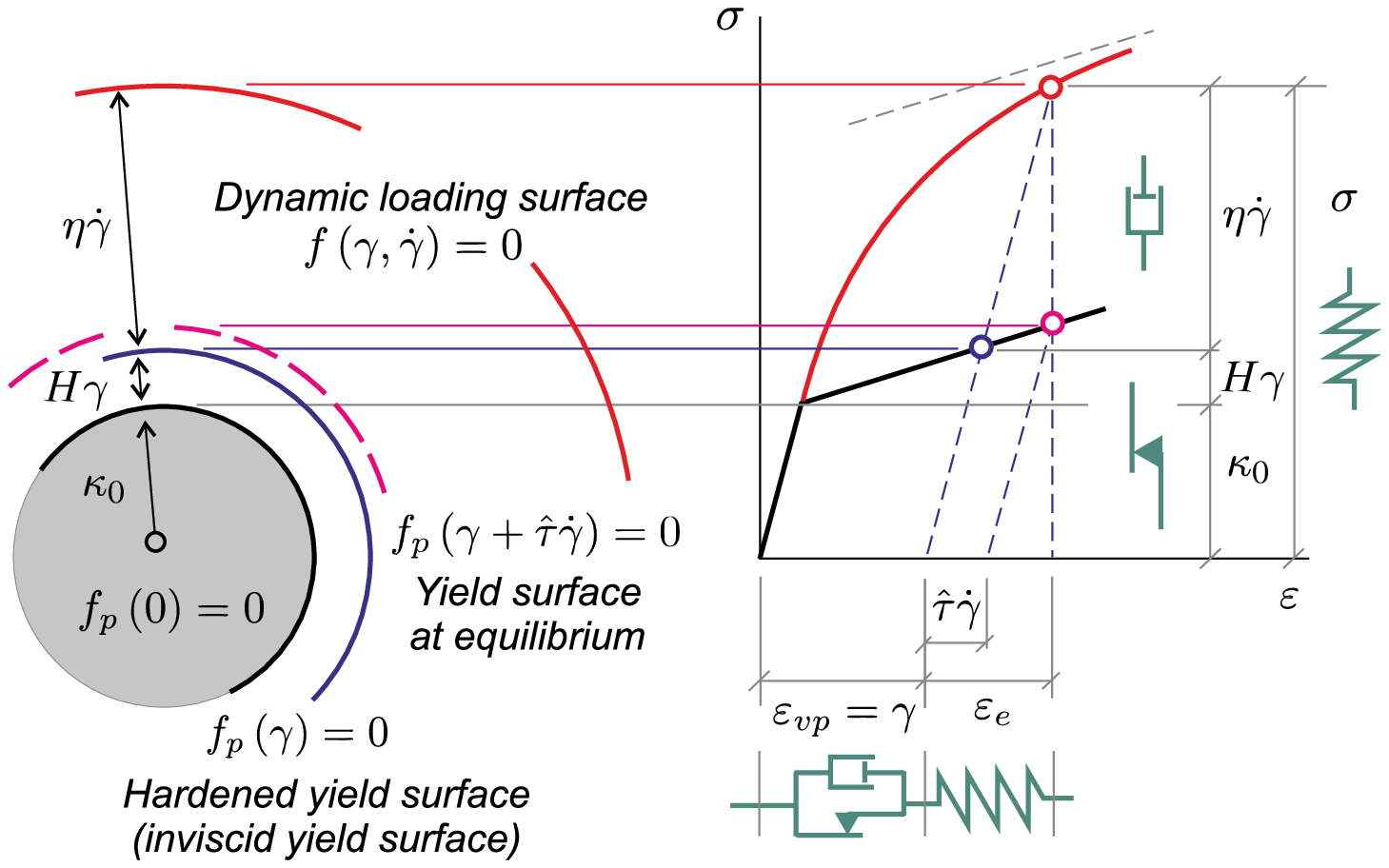}\caption{Surfaces
involved in the viscoplastic theory considering the Bingham model, and their
relation with the uniaxial loading curve and related quantities. Function $f$
interpreted as a ``dynamic loading function''. Function $f_{p}$ is interpreted
as the boundary of the elastic domain (yield surface). The viscoplastic strain
increment $\hat\tau\dot\gamma$ during relaxation produces an extra hardening}%
\label{fig.yield_surface}%
\end{figure}

$f_{p}(\boldsymbol{\varepsilon}_{e},\gamma)$ in Eq. (\ref{fpeq}) is the
classical plasticity (inviscid) criterion and $f\left(
\boldsymbol{\varepsilon}_{e},\gamma,\dot{\gamma}\right) =0 $ can be
interpreted as a ``dynamic loading surface'', which changes during the
deformation process by work-hardening effects and by the influence of the
strain-rate effect, as shown in Figure \ref{fig.yield_surface}.

\subsection{Continuum theory}

The previous equation implies that during the continuum flow with $\dot
{\gamma}>0$ we must also have $\dot{f}=0$ regardless of the value of the other
variables, so the requirement $f=0$ is maintained with, for example, changes
in the stress (as long as the condition $\dot{\gamma}>0$ still holds). Then,
considering any possible change in the variables:%
\begin{equation}
\dot{f}\left(  \boldsymbol{\varepsilon}_{e},\gamma,\dot{\gamma}\right)
=\frac{\partial f}{\partial\boldsymbol{\varepsilon}_{e}}:\left(
^{tr}\boldsymbol{\dot{\varepsilon}}_{e}+~^{ct}\boldsymbol{\dot{\varepsilon}%
}_{e}\right)  +\frac{\partial f}{\partial\gamma}\dot{\gamma}+\frac{\partial
f}{\partial\dot{\gamma}}\ddot{\gamma}=0 \label{fdot}%
\end{equation}
or%
\begin{equation}
\dot{f}=\underset{%
\begin{array}
[c]{c}%
^{tr}\dot{f}\equiv\left.  \dot{f}\right\vert _{D^{p}=0}\equiv\left.  \dot
{f}\right\vert _{\dot{\gamma}=0,\ddot{\gamma}=0}%
\end{array}
}{\underbrace{\frac{\partial f}{\partial\boldsymbol{\varepsilon}_{e}}%
:~^{tr}\boldsymbol{\dot{\varepsilon}}_{e}}}+\underset{%
\begin{array}
[c]{c}%
^{ct}\dot{f}\equiv\left.  \dot{f}\right\vert _{\mathcal{P}=0}\equiv\left.
\dot{f}\right\vert _{\boldsymbol{\dot{\varepsilon}}=0}%
\end{array}
}{\underbrace{\frac{\partial f}{\partial\boldsymbol{\varepsilon}_{e}}%
:~^{ct}\boldsymbol{\dot{\varepsilon}}_{e}+\frac{\partial f}{\partial\gamma
}\dot{\gamma}+\frac{\partial f}{\partial\dot{\gamma}}\ddot{\gamma}}}=0
\end{equation}
which in this case using $~^{ct}\boldsymbol{\dot{\varepsilon}}_{e}%
=-c\dot{\gamma}\boldsymbol{\hat{n}}$ and $^{tr}\boldsymbol{\dot{\varepsilon}%
}_{e}=\boldsymbol{\dot{\varepsilon}}$, and denoting $\kappa^{\prime}%
=d\kappa/d\gamma$ and $g^{\prime}=dg/d\dot{\gamma}$, is%
\begin{equation}
\dot{f}=\underset{%
\begin{array}
[c]{c}%
^{tr}\dot{f}%
\end{array}
}{\underbrace{c~\overset{%
\begin{array}
[c]{c}%
\boldsymbol{\hat{n}}:\boldsymbol{\dot{\sigma}}%
\end{array}
}{\overbrace{\boldsymbol{\hat{n}}:\mathbb{C}_{e}:\dot{\boldsymbol{\varepsilon
}}}}}}~\underset{%
\begin{array}
[c]{c}%
^{ct}\dot{f}%
\end{array}
}{\underbrace{-(c^{2}\boldsymbol{\hat{n}}:\mathbb{C}_{e}:\boldsymbol{\hat{n}%
}+\kappa^{\prime})\dot{\gamma}-g^{\prime}\ddot{\gamma}}}=0 \label{fdot2}%
\end{equation}
with
$\ddot{\gamma}=d\dot{\gamma}/dt$. The solution of this differential equation
gives the value of $\dot{\gamma}$ that maintains the thermodynamic
consistency, i.e. $f=0$ regardless of the changes in stress. By defining
---note the definition for the hardened case
\begin{equation}
\hat{\tau}:=\dfrac{g^{\prime}(\dot{\gamma})}{c^{2}\boldsymbol{\hat{n}%
}:\mathbb{C}_{e}:\boldsymbol{\hat{n}}+\kappa^{\prime}(\gamma)}\text{
(relaxation time)}%
\end{equation}
and
\begin{equation}
\dot{\gamma}_{\infty}:=\dfrac{\boldsymbol{\hat{n}}:\mathbb{C}_{e}%
:c\boldsymbol{\dot{\varepsilon}}}{c^{2}\boldsymbol{\hat{n}}:\mathbb{C}%
_{e}:\boldsymbol{\hat{n}}+\kappa^{\prime}(\gamma)}\text{ (inviscid rate)}
\label{inviscidrate}%
\end{equation}
the previous Eq. \eqref{fdot2} can be re-written as
\begin{equation}
\ddot{\gamma}+\frac{\dot{\gamma}}{\hat{\tau}}=\frac{\dot{\gamma}_{\infty}%
}{\hat{\tau}} \label{ODE2_v1}%
\end{equation}
and taking $\Gamma:=\dot{\gamma}$ for now for notational convenience, the
previous equation leads to a first-order \emph{scalar} differential equation
in $\Gamma$:%

\begin{equation}
\dot{\Gamma}+\frac{\Gamma}{\hat{\tau}}=\frac{\dot{\gamma}_{\infty}}{\hat{\tau
}} \label{ODE2}%
\end{equation}

Depending on the material parameters, the above equation could be a linear
differential equation or a nonlinear differential equation. For developing the
main ideas, we hereby particularize to the quite typical case in which
$\kappa^{\prime}=H$ and $g^{\prime}=\eta$ are constant (e.g. a linear
hardening and a constant viscosity $\eta$). Then, the solution of Eq.
\eqref{ODE2} can be determined as follows, if we assume that also
$c\boldsymbol{\hat{n}}:\mathbb{C}_{e}:\dot{\boldsymbol{\varepsilon}}$ is
constant (constant speed test, the case relevant for the incremental
formulation below)$\allowbreak$%

\begin{equation}
\left\{
\begin{array}
[c]{l}%
\Gamma\equiv\dot{\gamma}=\dot{\gamma}_{\infty}=\dfrac{c\boldsymbol{\hat{n}%
}:\mathbb{C}_{e}:\dot{\boldsymbol{\varepsilon}}}{c^{2}\boldsymbol{\hat{n}%
}:\mathbb{C}_{e}:\boldsymbol{\hat{n}}+H}>0\text{ \ if }\eta=0\medskip\\
\Gamma\equiv\dot{\gamma}=\dot{\gamma}_{\infty}+\bar{C}\exp\left(  -\dfrac
{t}{\hat{\tau}}\right)  >0\text{ \ if }\eta\neq0
\end{array}
\right.  \label{gammadot}%
\end{equation}
where $\bar{C}$ is a constant determined by $\dot{\gamma}\left(
t=t_{0}\right)  =:\dot{\gamma}_{0}$ as $\bar{C}=\left[  \dot{\gamma}_{0}%
-\dot{\gamma}_{\infty}\right]  \exp\left(  {t_{0}}/{\hat{\tau}}\right)  $, so
the second Equation (\ref{gammadot}) is%

\begin{equation}
\dot{\gamma} = \underset{
\begin{array}
[c]{c}%
\text{\textquotedblleft equilibrated\textquotedblright}\\
\text{(i.e. at }t\rightarrow\infty\text{)}\\
\text{or inviscid}%
\end{array}
} {\underbrace{\dot{\gamma}_{\infty}}}+\underset{%
\begin{array}
[c]{c}%
\text{\textquotedblleft non-equilibrium\textquotedblright\ rate}\\
\text{(viscous contribution)}%
\end{array}
} {\underbrace{\overset{%
\begin{array}
[c]{c}%
\text{\textquotedblleft non-equilibrium\textquotedblright}\\
\text{forcing rate}%
\end{array}
}{\overbrace{\left[  \dot{\gamma}_{0}-\dot{\gamma}_{\infty}\right]  }}%
\exp\left(  -\dfrac{t-t_{0}}{\hat{\tau}}\right)  }} \label{dotgamma}%
\end{equation}
in which we can interpret that $\dot{\gamma}_{neq}:=\dot{\gamma}_{0}%
-\dot{\gamma}_{\infty}$ is the non-equilibrated rate and $\dot{\gamma}%
_{\infty}$ corresponds to the elastoplastic (inviscid) rate solution, i.e. the
solution with $\eta\rightarrow0$ or at $t\rightarrow\infty$. Another physical
interpretation typical of visco\textit{elasticity} is obtained rearranging
terms%
\begin{equation}
\dot{\gamma}(t)=\underset{%
\begin{array}
[c]{c}%
\text{I.C. vanishing term}%
\end{array}
}{\underbrace{\dot{\gamma}_{0}\exp\left(  -\dfrac{t-t_{0}}{\hat{\tau}}\right)
}}+\underset{%
\begin{array}
[c]{c}%
\text{Steady-state accommodating term}%
\end{array}
}{\underbrace{\dot{\gamma}_{\infty}~\left[  1-\exp\left(  -\dfrac{t-t_{0}%
}{\hat{\tau}}\right)  \right]  }}\geq0 \label{eq29}%
\end{equation}
i.e. the first addend is the influence of the initial condition $\dot{\gamma
}_{0}$ vanishing in time, and the second term is the steady-state term
$\dot{\gamma}_{\infty}$ being enforced in time. Substitute Eq.
\eqref{dotgamma} in Eq. \eqref{ODE2_v1} to get the speed at which this
adaptation process takes place---namely the speed at which $\dot{\gamma}%
_{neq}$ is cancelled-out
\begin{equation}
\ddot{\gamma}(t)=-\frac{\dot\gamma_{neq}
}{\hat{\tau}}\exp\left(  -\dfrac{t-t_{0}}{\hat{\tau}}\right)
\label{gamma2dot}%
\end{equation}
Note that Eq. (\ref{gamma2dot}) is in essence similar to the Perzyna model but
in second derivative and fully written in kinematic quantities, in rate form(consider that at $t=t_0$ we have $\ddot\gamma=-\dot\gamma_{neq}/\hat\tau$). Of course in the
continuum theory, the incremental consistency parameter is obtained by
integration of Eq. (\ref{dotgamma}) from $t=t_{0}$ to a time $t$ as%
\begin{align}
\gamma &  =\gamma_{0}+\int_{t=t_{0}}^{t}\left[  \dot{\gamma}_{\infty}%
+\dot{\gamma}_{neq}\exp\left(  -\frac{t-t_{0}}{\hat{\tau}}\right)  \right]
dt\nonumber\\
&  =\gamma_{0}+\dot{\gamma}_{\infty}\left(  t-t_{0}\right)  +\hat{\tau}\left(
\dot{\gamma}_{0}-\dot{\gamma}_{\infty}\right)  \left[  1-\exp\left(
-\frac{t-t_{0}}{\hat{\tau}}\right)  \right]  \label{eq32}%
\end{align}
where%
\begin{equation}
\int_{t_{0}}^{t}\exp\left(  -\dfrac{t-t_{0}}{\hat{\tau}}\right)  dt=-\hat
{\tau}\left.  \exp\left(  -\dfrac{t-t_{0}}{\hat{\tau}}\right)  \right\vert
_{t_{0}}^{t}=\hat{\tau}\left[  1-\exp\left(  -\dfrac{t-t_{0}}{\hat{\tau}%
}\right)  \right]  =:\hat{\tau}\xi\left(  t-t_{0}\right)
\end{equation}
is a result that we will use repeatedly below with $\xi(t-t_{0})$ as compact notation.
A relevant case is when a sudden relaxation takes place. In this case, taking
$t_{0}=0$, $\gamma_{0}\neq0$, $\dot\gamma_{0}\neq0$, $\dot\gamma_{\infty}= 0$,
$t\rightarrow\infty$, we get the value at equilibrium, namely $\gamma=
\gamma_{0} + \hat\tau\dot\gamma_{0} =:\gamma^{r}_{\infty}$; i.e. the
equilibrium viscoplastic strain $\gamma_{\infty}^{r}$ is $\hat\tau\dot
\gamma_{0}$ away from $\gamma_{0}$.

In the viscoplastic case, we do not use any unloading/reloading condition as
in plasticity. However, there are two similar cases: conservative case and
dissipative case. The condition for conservative case is simply physically
determined by $\dot{\gamma}=0$, in which case we may have $f\neq0$ and
$\dot{f}\neq0$, but also $f=0$. The condition for dissipative case simply
requires $\dot{\gamma}>0$, which implies that $f=\dot{f}=0$ by the first
principle. The case $\dot{\gamma}<0$ is not possible by definition (would
entail a negative dissipation, violating the second law of thermodynamics).
Both conservative and dissipative cases are distinguished by $\dot{\gamma}$,
not by $f$; i.e. it is $\dot{\gamma}$, computed from its own evolution
Equation (\ref{dotgamma}) the quantity to check, and its numerical integration
must just guarantee that $f\ngtr0$.
However, the start of viscoplastic loading from elastic one is detected by $f_p>0$.

In order to obtain the continuous viscoplastic tangent moduli tensor, we can
use the constitutive equation in the rate form along Eq. (\ref{eq29})
\begin{align}
\dot{\boldsymbol{\sigma}}  &  =\mathbb{C}:\dot{\boldsymbol{\varepsilon}%
}=\mathbb{C}_{e}:\dot{\boldsymbol{\varepsilon}}_{e}=\mathbb{C}_{e}:(~^{tr}%
\dot{\boldsymbol{\varepsilon}}_{e}+~^{ct}\dot{\boldsymbol{\varepsilon}}%
_{e})=\mathbb{C}_{e}:(\dot{\boldsymbol{\varepsilon}}-c\dot{\gamma
}\boldsymbol{\hat{n}})\nonumber\\
&  =\left[  \mathbb{C}_{e}-\xi(t-t_{0})\dfrac{(\mathbb{C}_{e}:\boldsymbol{\hat
{n}})\otimes(\mathbb{C}_{e}:\boldsymbol{\hat{n}})}{\boldsymbol{\hat{n}%
}:\mathbb{C}_{e}:\boldsymbol{\hat{n}}+\kappa^{\prime}/c^{2}}\right]
:\dot{\boldsymbol{\varepsilon}}-c\dot{\gamma}_{0}\exp\left(  -\dfrac{t-t_{0}%
}{\hat{\tau}}\right)  \mathbb{C}_{e}:\boldsymbol{\hat{n}} \label{rateform}%
\end{align}
If the initial condition is $\dot{\gamma}_{0}=0$, the last addend vanishes, so
Eq. \eqref{rateform} is
\begin{equation}
\dot{\boldsymbol{\sigma}}=\underset{%
\begin{array}
[c]{c}%
\mathbb{C}%
\end{array}
}{\underbrace{\left[  \mathbb{C}_{e}-\dfrac{\mathbb{C}_{e}:\boldsymbol{\hat
{n}}\otimes\mathbb{C}_{e}:\boldsymbol{\hat{n}}}{\boldsymbol{\hat{n}%
}:\mathbb{C}_{e}:\boldsymbol{\hat{n}}+\kappa^{\prime}/c^{2}}\xi(t-t_{0}%
)\right]  }}:\dot{\boldsymbol{\varepsilon}}%
\end{equation}
where $\mathbb{C}$ denotes the continuous viscoplastic tangent modulus tensor.
Note that $\mathbb{C}$ is bounded by the elastic tangent modulus tensor for
the instantaneous response ($t=t_{0}$ and $\xi(t_{0})=0$), and by the
elastoplastic tangent modulus tensor for the long term response ($t\rightarrow
\infty$ and $\xi(\infty)\rightarrow1$), that is%

\begin{equation}
\mathbb{C}=%
\begin{cases}
\mathbb{C}_{e} & \quad\text{for}\quad t\rightarrow t_{0}\text{ or }%
\eta\rightarrow\infty\\
\mathbb{C}_{e}-\dfrac{\mathbb{C}_{e}:\boldsymbol{\hat{n}}\otimes\mathbb{C}%
_{e}:\boldsymbol{\hat{n}}}{\boldsymbol{\hat{n}}:\mathbb{C}_{e}%
:\boldsymbol{\hat{n}}+\kappa^{\prime}/c^{2}}\equiv\mathbb{C}_{ep} &
\quad\text{for}\quad t\rightarrow\infty\text{ or }\eta\rightarrow0
\end{cases}
\label{Tangent limites}%
\end{equation}

Obviously, in the cases in which the coefficients of the differential equation
are not constant, the solution depends on those functions, but the previous
expressions may be considered as an approximation if that nonlinearity is weak
or the computational steps, small. A general algorithmic solution, including
nonlinear functions, is given below.

\subsection{Proportional loading cases}

Several monotonic, uniaxial cases are of interest to understand the behaviour
of the model, so they are briefly discussed here for the linear case.

\begin{figure}[pth]
\centering
\includegraphics[width=1.0\textwidth]{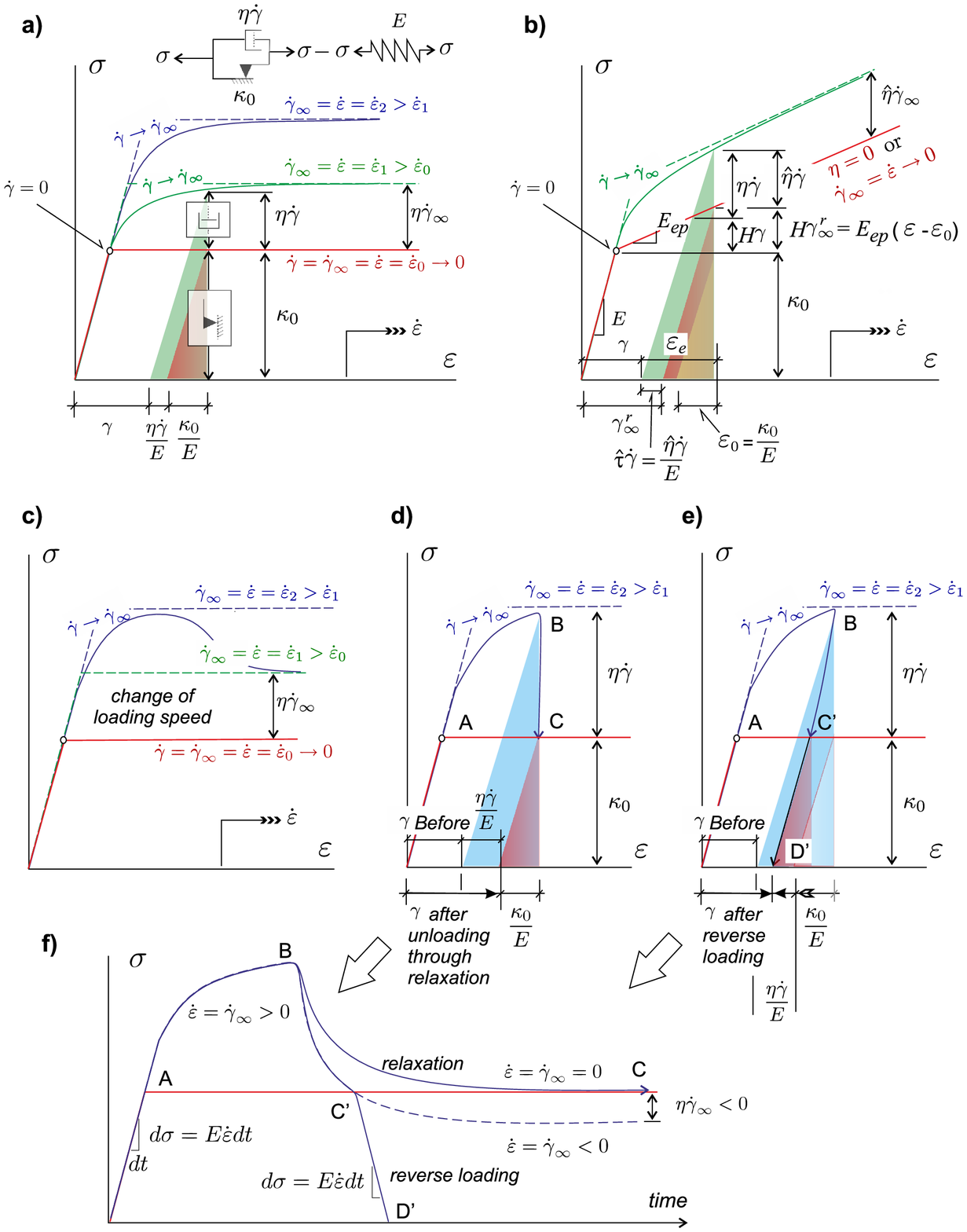}\caption{Uniaxial
proportional loading. a) Monotonic loading at constant speed with no
hardening. b) Monotonic loading at constant speed with hardening. c) Change of
speed during monotonic loading. d) Viscous relaxation. e) Reverse
loading/unloading. f) Stress versus time in relaxation and reverse-loading
cases. }%
\label{dibujos}%
\end{figure}

\subsubsection{Constant rate loading case}

The first case is when $\dot{\varepsilon}$ is a constant uniaxial loading (i.e. 1D). In
this case, until $f_{p}>0$ at $t>t_{0}$, the stress rate is $\dot{\sigma
}=E\dot{\varepsilon}$, where $E$ is the Young modulus and $\dot{\sigma}$ is
the uniaxial stress rate. Once $f_{p}>0$, if there is no hardening
($\kappa^{\prime}=H=0$), since $\dot{\gamma}_{0}=0$ at $t_{0}$ (onset of
viscoplastic loading), the 1D version of Eq. (\ref{rateform}) results in%
\begin{equation}
\dot{\sigma}=E\dot{\varepsilon}-E\dot{\varepsilon}\left[  1-\exp\left(
-\dfrac{t-t_{0}}{\hat{\tau}}\right)  \right]
\end{equation}
with $\hat{\tau}=\eta/E$. Note that for $t=t_{0}$ we have the elastic
$\dot{\sigma}=E\dot{\varepsilon}$ and for $t\rightarrow\infty$ we have
$\dot{\sigma}=0$, which is the rate of the perfect plasticity solution. The
integral from $t=t_{0}$ to $t$ is%
\begin{align}
\left.  \sigma-\kappa_{0}=\int_{t_{0}}^{t}\dot{\sigma}dt\right.   &
=\hat{\tau}E\dot{\varepsilon}\xi(t-t_{0})=\eta\dot{\gamma}_{\infty}\xi
(t-t_{0})
\end{align}
giving the limits $\sigma=\kappa_{0}$ for $t=t_{0}$ and $\sigma=\kappa
_{0}+\eta\dot{\gamma}_{\infty}$ for $t\rightarrow\infty$. This is shown in
Fig. \ref{dibujos}a. The effect in this figure of increasing $\eta$ is the
same as increasing the rate $\dot{\varepsilon}=\dot{\gamma}_{\infty}$.

For the case of simple shear, the shear stress $\sigma_{s}$ is computed in
terms of the tensorial shear strain $\varepsilon_{s}$ (half the engineering
one) from the 3D solution, Eq. (\ref{rateform}), as%
\begin{align}
\left.  \sigma_{s}-\tfrac{1}{\sqrt{3}}\kappa_{0}=\int_{t_{0}}^{t}\dot{\sigma
}_{s}dt\right.  =\hat{\tau}2\mu\dot{\varepsilon}_{s}\xi(t-t_{0})=\frac{\eta
}{c^{2}}\dot{\varepsilon}_{s}\xi(t-t_{0})
\end{align}
where we used $\dot{\boldsymbol{\varepsilon}}=\dot{\varepsilon}_{s}%
\boldsymbol{\hat{n}}$ and $\mathbb{C}_{e}:\boldsymbol{\hat{n}}$ $=2\mu
\boldsymbol{\hat{n}}$ and $\boldsymbol{\hat{n}}:\mathbb{C}_{e}%
:\boldsymbol{\hat{n}}$ $=2\mu$ and $\dot{\gamma}_{\infty}=c2\mu\dot
{\varepsilon}_{s}/\left(  2\mu c^{2}\right)  =\dot{\varepsilon}_{s}/c$ and
$\hat{\tau}=\eta/\left(  2\mu c^{2}\right)  $. Then $\sigma_{s}=\kappa
_{0}/\sqrt{3}+\left(  \eta/c^{2}\right)  \dot{\varepsilon}_{s}\xi(t-t_{0})$.

The case with linear hardening shown in Fig. \ref{dibujos}b is similar. Recall
the definition of the relaxed viscoplastic strain $\gamma_{\infty}^{r}%
:=\gamma+\hat{\tau}\dot{\gamma} \label{ginfr}$ where the first addend is the
current viscoplastic strain and the second one is its potential increment if a
sudden relaxation process takes place. Define $\hat{\eta}:=\hat{\tau}E$, so
$\hat{\eta}=\left[  E/\left(  E+H\right)  \right]  \eta$, i.e. $\eta=\hat
{\eta}$ if no hardening is present. Recall that $\dot{\gamma}=\dot{\gamma
}_{\infty}\xi\left(  t-t_{0}\right)  $ with $\dot{\gamma}_{\infty}=\left(
E\dot{\varepsilon}\right)  /\left(  E+H\right)  $ for $\dot{\gamma}_{0}=0$.
Then
\begin{align}
\sigma-\kappa_{0}  &  =E_{ep}\left[  \varepsilon\left(  t\right)
-\varepsilon_{0}\right]  +\eta_{u-eq}\dot{\varepsilon}\xi\left(
t-t_{0}\right) \\
&  =E_{ep}\left[  \varepsilon\left(  t\right)  -\varepsilon_{0}\right]
+\hat{\eta}\dot{\gamma}_{\infty}\xi\left(  t-t_{0}\right) \nonumber\\
&  =H\gamma_{\infty}^{r}+\hat{\eta}\dot{\gamma}\left(  t\right)
\;\;\;=H\gamma+\eta\dot{\gamma}\left(  t\right)\label{eq39}
\end{align}
so $\sigma=\kappa_{0}+H\gamma_{\infty}^{r}+\hat{\eta}\dot{\gamma}\left(
t\right)  =\kappa_{0}+H\gamma+\eta\dot{\gamma}\left(  t\right)  $, and where
we defined an equivalent uniaxial viscosity for the hardening case as
$\eta_{u-eq}=\eta E^{2}/\left(  E+H\right)  ^{2}=\hat{\eta}E/\left(
E+H\right)  $ and $E_{ep}=HE/(E+H)$. In Eq. (\ref{eq39}), the first identity is written in terms of yield stress at
equilibrium ($\kappa_{0}+H\gamma_{\infty}^{r}$), see Fig.
\ref{fig.yield_surface}, whereas the second one is written in terms of the
current one ($\kappa_{0}+H\gamma$). Note that
\begin{equation}
\frac{d\sigma}{d\varepsilon}\left(  t\rightarrow\infty\right)  =E_{ep}%
\end{equation}
as expected from Eq. (\ref{Tangent limites}). The sketch in Fig.
\ref{dibujos}b is better interpreted in terms of the quantities at
equilibrium. Note that during relaxation with $\dot{\varepsilon}=0$ we have
$\gamma\rightarrow\gamma_{\infty}^{r}$, and $\varepsilon_{e}$ decreases by
$\left(  \hat{\eta}/E\right)  \dot{\gamma}=\hat{\tau}\dot{\gamma}$. Here
$\hat{\eta}$ compensates for the hardening, because part of the stress in the
dashpot $\eta\dot{\gamma}$ will be absorbed by hardening. Then, the
stress-strain curve is initially the same as the elastic one $E$, until
$\sigma=\kappa_{0}$. Thereafter it will exponentially adapt to a line with
slope $E_{ep}$, but shifted a constant $\hat{\eta}\dot{\gamma}_{\infty}%
\equiv\eta_{u-eq}\dot{\varepsilon}$ from the elastoplastic one. \newline

The case of simple shear may be obtained again directly from the 3D case
taking $\dot{\boldsymbol{\varepsilon}}=\dot{\varepsilon}_{s}\boldsymbol{\hat
{n}}$ and $\mathbb{C}_{e}:\boldsymbol{\hat{n}}$ $=2\mu\boldsymbol{\hat{n}}$
and $\boldsymbol{\hat{n}}:\mathbb{C}_{e}:\boldsymbol{\hat{n}}$ $=2\mu$ and
$\hat{\tau}=\eta/\left(  2\mu c^{2}+H\right)  $. In this case%
\begin{equation}
\sigma_{s}-\frac{1}{\sqrt{3}}\kappa_{0}=\int_{t_{0}}^{t}\dot{\sigma}_{s}dt
=2\mu_{ep}\left[  \varepsilon_{s}\left(  t\right)  -\varepsilon_{0s}\right]
+\frac{\eta_{s-eq}}{c^{2}}\dot{\varepsilon}_{s}\xi\left(  t-t_{0}\right)
\end{equation}
where we defined $\eta_{s-eq}:=\eta\left(  2\mu\right)  ^{2}/ \left(
2\mu+H/c^{2}\right)  ^{2} $ and $2\mu_{ep}:=2\mu H/(2\mu c^{2}+H)$. Note that
for $H=0$ we have $\eta_{s-eq}=\eta$, and that, also as expected from Eq.
(\ref{Tangent limites})
\begin{equation}
\frac{d\sigma}{d\varepsilon_{s}}\left(  t\rightarrow\infty\right)  =2\mu_{ep}%
\end{equation}
Then, the shear stress versus shear strain (tensorial) has an initial
(elastic) slope of $2\mu$ until $\sigma_{s}=\kappa_{0}/\sqrt{3}$ and a
limiting slope of $2\mu_{ep}$ for $t\rightarrow\infty$; and an offset from the
hardened elastoplastic line of $\eta_{s-eq}\dot{\varepsilon}_{s}$ enforced
progressively through the exponential-type function $\xi\left(  t-t_{0}%
\right)  $.

\subsubsection{Change of speed}

If there is a {change of speed}, the stress path simply changes the horizontal
asymptote, as shown in Fig. \ref{dibujos}c, because $\dot{\gamma}_{\infty}$
also does.

\subsubsection{Relaxation, unloading and reverse loading}

In the case of sudden stop in strain loading, i.e. $\dot{\varepsilon}=0$, a
{relaxation process} occurs to the inviscid as shown in Fig. \ref{dibujos}d
and in Fig. \ref{dibujos}f in time. The unloading curve in this latter case is%
\begin{equation}
\sigma=\kappa_{0}+\eta\dot{\gamma}=\kappa_{0}+\eta\dot{\gamma}_{0}\exp\left(
-\dfrac{t-t_{0}}{\hat{\tau}}\right)
\end{equation}
where $\dot{\gamma}_{0}$ is the value at the beginning of the relaxation
process and $t_{0}$ is the instant at which relaxation begun. The tangent of
the relaxation in time is%
\begin{equation}
\frac{d\sigma}{dt}=-E\dot{\gamma}_{0}\exp\left(  -\dfrac{t-t_{0}}{\hat{\tau}%
}\right)
\end{equation}
In the case of hardening $\sigma=\kappa_{0}+H\gamma+\eta\dot{\gamma}$ until
$\dot{\gamma}=0$. Since $\dot{\varepsilon}=0$ we have $\dot{\gamma}_{\infty
}=0$, so $\dot{\gamma}(t)=\dot{\gamma}_{0}\exp\left[  -\left(  t-t_{0}\right)
/\hat{\tau}\right]  $ and $\gamma=\gamma_{0}+\hat{\tau}\dot{\gamma}_{0}%
-\hat{\tau}\dot{\gamma}_{0}\exp\left[  -\left(  t-t_{0}\right)  /\hat{\tau
}\right]  $. Then
\begin{equation}
\sigma=\kappa_{0}+H\left(  \gamma_{0}+\hat{\tau}\dot{\gamma}_{0}-\hat{\tau
}\dot{\gamma}_{0}\exp\left[  -\left(  t-t_{0}\right)  /\hat{\tau}\right]
\right)  +\eta\dot{\gamma}_{0}\exp\left(  -\dfrac{t-t_{0}}{\hat{\tau}}\right)
\end{equation}
and for $t\rightarrow\infty$ we get $\sigma=\kappa_{0}+H\left(  \gamma
_{0}+\hat{\tau}\dot{\gamma}_{0}\right)  $, as it should be expected from Eq.
(\ref{ginfr}).

A similar process occurs if there is a {reverse loading or unloading}, as
shown in Fig. \ref{dibujos}d and in Fig. \ref{dibujos}f in time. Consider a
change from a positive (loading) $\dot{\varepsilon}=\dot{\varepsilon}_{l}>0$
to a negative (unloading) rate $\varepsilon=\dot{\varepsilon}_{u}<0$. Then
$\dot{\gamma}_{\infty}\left(  \dot{\varepsilon}_{u}\right)  <0$. The stress is%
\begin{equation}
\sigma=\underset{%
\begin{array}
[c]{c}%
<\kappa_{0}%
\end{array}
}{\underbrace{\kappa_{0}+\eta\dot{\gamma}_{\infty}}}+\underset{%
\begin{array}
[c]{c}%
>0
\end{array}
}{\underbrace{\eta\left(  \dot{\gamma}_{0}-\dot{\gamma}_{\infty}\right)  }%
}\underset{%
\begin{array}
[c]{c}%
1\rightarrow0
\end{array}
}{\underbrace{\exp\left(  -\dfrac{t-t_{0}}{\hat{\tau}}\right)  }}%
\end{equation}
The stress relaxes towards an horizontal asymptote at $\kappa_{0}+\eta
\dot{\gamma}_{\infty}<\kappa_{0}$, with a speed in time given by%
\begin{equation}
\frac{d\sigma}{dt}=-E\left(  \dot{\gamma}_{0}-\dot{\gamma}_{\infty}\right)
\exp\left(  -\dfrac{t-t_{0}}{\hat{\tau}}\right)
\end{equation}
This viscous relaxation takes place until the inviscid yield surface $f_{p}=0$
is crossed (i.e. when $\sigma=\kappa_{0}$), which happens at time%
\begin{equation}
t=t_{0}-\hat{\tau}\log\left(  \frac{\dot{\gamma}_{\infty}}{ \dot{\gamma
}_{\infty}-\dot{\gamma}_{0} }\right)
\end{equation}
which obviously gives the limit $t\rightarrow t_{0}+\infty$ for $\dot{\gamma
}_{\infty}\rightarrow0$, corresponding to the relaxation case. After $f_{p}=0$
is crossed, the unloading continues elastically, see Fig. \ref{dibujos}d. Note
that even in the reverse loading case, $\gamma$ continues to increase until
the inviscid yield surface is crossed; i.e. as long as $f_{p}>f=0$.

\begin{figure}[h]
\centering
\includegraphics[width=0.5\textwidth]{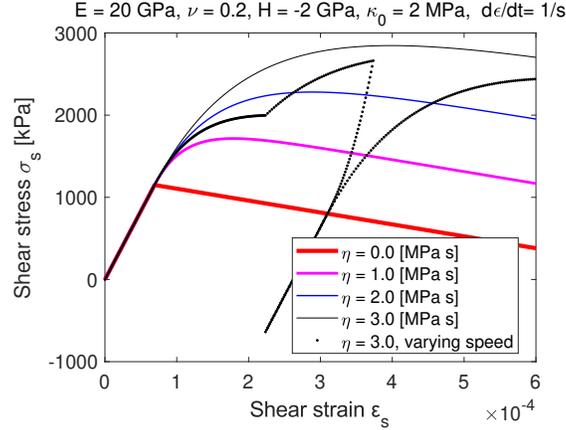}\caption{Influence
of the loading rate or the viscosity parameter $\eta$ on the predictions of
the model. $\eta=0$ corresponds to the inviscid solution. Note that a change
of loading rate has the same influence as a change of viscosity. Note also
that the limiting elastoplastic tangent is reached in all cases, i.e.
$2\mu_{eq} = 2\mu H/(2\mu c^{2} + H)=-1.45$ MPa, and that the stress offset in
the asymptote is given by $\eta_{s-eq}\dot\varepsilon_{s}/c^{2}$. Loading rate
of continuous curves: $\dot\varepsilon=1/s$. Dotted curve has sharp changes in
loading rates, being those of $\dot\varepsilon_{s}=0.5/s$, $1.0/s$, $-0.5/s$,
$1.0/s$. Note that dissipative behaviour still takes place during reverse
loading until the plastic yield function (elastic domain) is reached.}%
\label{fig.vonMises}%
\end{figure}

In Fig. \ref{fig.vonMises} we show the behavior of the model under simple
shear for different viscosities and a softening modulus, where the previous
effects may be observed. 

\section{Incremental theory of $J_{2}$--viscoplasticity with linear isotropic
hardening}

We develop an incremental solution for a step to build the computational
implicit algorithm, first in this section with attention to the linear case.
The solution of the step depends on whether the step is fully elastic (which
solution is trivial), fully viscoplastic, or mixed elastic-to-viscoplastic or
viscoplastic-to-elastic. We denote the time step by left-superindices as in
$~^{t}(\bullet)$, following the notation in e.g.
\cite{Bathe96,KojicBathe,Dvorkin}.

\subsection{All the step is dissipative}

\label{subsec.31}

In the typical predictor-corrector algorithms, the two components
$^{tr}\boldsymbol{\dot{\varepsilon}}_{e}$ and $^{ct}\boldsymbol{\dot
{\varepsilon}}_{e}$ are integrated in two successive sub-steps; indeed the
$-^{ct}\boldsymbol{\dot{\varepsilon}}_{e}$ is identified as $\boldsymbol{\dot
{\varepsilon}}_{p}$ in classical procedures, an identification which only
holds at small strains \cite{LatMonPlas} and which allows for the
identification $-\Delta^{ct}\boldsymbol{\varepsilon}_{e}\equiv\Delta\boldsymbol{\varepsilon}_{p}$. The
first of them (the trial part) is purely hyperelastic, conservative, i.e.
during the step $^{tr}\mathcal{D}^{p}=0$, but changes the stored energy from
$^{t}\Psi:=\Psi\left(  ^{t}\boldsymbol{\varepsilon}_{e}\right)  $ to
$^{tr}\Psi:=\Psi\left(  ^{tr}\boldsymbol{\varepsilon}_{e}\right)  $. Then, the
error in the fulfillment of the first principle during a step comes only from
the dissipative part in a subsequent substep. Unfortunately, whereas this type
of predictor-corrector algorithms are well-suited for elastoplasticity, in
viscoplasticity the predictor phase cannot be easily isolated from the
corrector phase because of the time-dependence (both effects occur
simultaneously). This is manifest by the comparison of both Equations
(\ref{gammadot}). The first one is independent of time so an increment may be
applied to both hand sides and the result is independent of the time in which
the increments took place, e.g. $\Delta\gamma=\dot{\gamma}\Delta t$ and
$\Delta\varepsilon=\dot{\varepsilon}\Delta t$ so $\Delta t$ cancels out.
However, in the second one, time cannot be eliminated because the speed at
which the increment takes place is important, since that speed changes the
dissipated energy through the dashpot; for example in a quasi-static
deformation the dashpot does not dissipate energy whatever the value of $\eta$
is, but in a very fast process most dissipation comes from the dashpot.
Noteworthy, Eq. (\ref{gammadot})$_{1}$ is equivalent to establish $^{t+\Delta
t}f_{p}=0$ integrated with a backward-Euler method (i.e. the solution from the
radial return algorithm of Wilkins \cite{Wilkins})%
\begin{equation}
\Delta\gamma=\frac{1}{c}\dfrac{2\mu\left(  ^{t+\Delta t}\boldsymbol{\hat{n}%
}:\Delta\boldsymbol{\varepsilon}\right)  }{2\mu+H/c^{2}}\equiv\frac{1}%
{c}\dfrac{2\mu\left(  ^{ tr}\boldsymbol{\hat{n}}:\Delta\boldsymbol{\varepsilon
}\right)  }{2\mu+H/c^{2}}=:\Delta\gamma_{\infty} \label{dgammainfty}%
\end{equation}
Then, considering still the linear case with constant $\kappa^{\prime}=H$,
$g^{\prime}=\eta$ and $\left(  \boldsymbol{\hat{n}}:\boldsymbol{\dot
{\varepsilon}}\right)  =\left(  \boldsymbol{\hat{n}}:\Delta
\boldsymbol{\varepsilon}\right)  /\Delta t$ during the step (so $\hat{\tau}$
and $\dot{\gamma}_{\infty}$ are also constant), the \emph{exact} integration
of the equivalent viscoplastic strain is (i.e. no error is introduced if
$\boldsymbol{\hat n}$ is constant, which happens in proportional loading)
\begin{align}
\Delta\gamma &  =\int_{t}^{t+\Delta t}\dot{\gamma}dt=\int_{t}^{t+\Delta
t}\left\{  \dot{\gamma}_{\infty}+\left[  ^{t}\dot{\gamma}-\dot{\gamma}%
_{\infty}\right]  \exp\left(  -\frac{\bar{t}-t}{\hat{\tau}}\right)  \right\}
d\bar{t}\nonumber\\
&  =~\dot{\gamma}_{\infty}\Delta t+\hat{\tau}\left[  ^{t}\dot{\gamma
}-~^{t+\Delta t}\dot{\gamma}_{\infty}\right]  \left[  1-\exp\left(
-\frac{\Delta t}{\hat{\tau}}\right)  \right]  \nless0 \label{delta gamma}%
\end{align}
with the definition given in Eq. (\ref{dgammainfty}) and the definition
\textit{during the current step} (i.e. from $t$ to $t+\Delta t$) of
$^{t+\Delta t}\dot{\gamma}_{\infty}\equiv~^{tr}\dot{\gamma}_{\infty}%
\equiv~\dot{\gamma}_{\infty}:=\Delta\gamma_{\infty}/\Delta t$. For small steps
we have $\left[  \left(  2\mu c^{2}+H\right)  /\eta\right]  \Delta t=:\Delta
t/\hat{\tau}\rightarrow0$, where the relaxation time in the present linear
case is
\begin{equation}
\hat{\tau}=\frac{\eta}{2\mu c^{2}+H}=\frac{\eta/c^{2}}{2\mu+H/c^{2}}%
\end{equation}
Note that the expected limits are recovered, e.g. $\Delta t/\hat{\tau}$ small
implies $\Delta\gamma\simeq$ $~^{t}\dot{\gamma}\Delta t$ and for $\Delta
t/\hat{\tau}\rightarrow\infty$ we have $\Delta\gamma\rightarrow\Delta
\gamma_{\infty}+\hat{\tau}~^{t}\dot{\gamma}$. If $^{t}\dot{\gamma}=0$ we
obtain $\Delta\gamma=\dot{\gamma}_{\infty}\left[  \Delta t-\hat{\tau}%
\xi(\Delta t)\right]  $. Consider Eq. \eqref{dotgamma} at $t+\Delta t$ where
the step has a uniform external strain speed given by $\boldsymbol{\dot
{\varepsilon}}=\Delta\boldsymbol{\varepsilon}/\Delta t$%
\begin{equation}
^{t+\Delta t}\dot{\gamma} =~^{t+\Delta t}\dot{\gamma}_{\infty}+\left[
~^{t}\dot{\gamma}-~^{t+\Delta t}\dot{\gamma}_{\infty}\right]  \exp\left(
-\frac{\Delta t}{\hat{\tau}}\right)  \nless0\label{gammadotn+1}%
\end{equation}
so%
\begin{equation}
\left.  \Delta\dot{\gamma}:=~^{t+\Delta t}\dot{\gamma}-~^{t}\dot{\gamma
}\right.  =\left(  ~^{t+\Delta t}\dot{\gamma}_{\infty}-~^{t}\dot{\gamma
}\right)  \left[  1-\exp\left(  -\frac{\Delta t}{\hat{\tau}}\right)  \right]
=~^{tr}\dot{\gamma}_{neq}~\xi(\Delta t) \label{gammadot2}%
\end{equation}
where $~^{tr}\dot{\gamma}_{neq}:=~^{t+\Delta t}\dot{\gamma}_{\infty}-~^{t}%
\dot{\gamma}$ is the trial non-equilibrated rate at $t+\Delta t$; i.e. the
difference between the \textquotedblleft at infinite\textquotedblright%
\ (inviscid) rate during the step $\Delta\gamma_{\infty}/\Delta t$ and the
actual one at the beginning of the step $~^{t}\dot{\gamma}$. The relaxation
case is obtained when $\boldsymbol{\dot{\varepsilon}}=\boldsymbol{0}$, i.e.
$\Delta\boldsymbol{\varepsilon}=\boldsymbol{0}.$ Then $\Delta\gamma_{\infty
}=\dot{\gamma}_{\infty}=0$ and $\Delta\gamma=~^{t}\dot{\gamma}~\hat{\tau}%
~\xi(\Delta t)$ and $\Delta\dot{\gamma}=-~^{t}\dot{\gamma}~\xi(\Delta t)$. In
such case,\ the zero rate $\dot{\gamma}=0$ is obtained with $\Delta\dot
{\gamma}=-~^{t}\dot{\gamma}$
\begin{equation}
-~^{t}\dot{\gamma}=-~^{t}\dot{\gamma}\left[  1-\exp\left(  -\frac{\Delta
t}{\hat{\tau}}\right)  \right]  \;\Rightarrow\;\frac{\Delta t}{\hat{\tau}%
}\rightarrow\infty\label{Stationary rate}%
\end{equation}
at time $\Delta t\rightarrow\infty$, where $\Delta\gamma=~^{t}\dot{\gamma
}~\hat{\tau}$---c.f. again Eq. (\ref{ginfr})

Consider the integration of the thermodynamical power balance (i.e. energy
balance) during the step using the previous relations%
\begin{equation}
\int_{t}^{t+\Delta t}~^{tr}\dot{f}dt=\left.  \Delta f\right\vert _{D^{p}%
=0}:=~^{tr}f-~^{t}f
\end{equation}%
\begin{equation}
\int_{t}^{t+\Delta t}~^{ct}\dot{f}dt=-\left(  2\mu c^{2}+H\right)
\Delta\gamma-\eta~\Delta\dot{\gamma}%
\end{equation}
Noteworthy, if we require energy conservation, so during the step $\Delta f=0$
(as to obtain $^{t+\Delta t}f=0$ if $^{t}f=0$), we have%
\begin{align}
^{t+\Delta t}f-~^{t}f  &  =\int_{t}^{t+\Delta t}~^{tr}\dot{f}dt+\int
_{t}^{t+\Delta t}~^{ct}\dot{f}dt\nonumber\\
&  =\left(  ~^{tr}f-~^{t}f\right)  -\left(  2\mu c^{2}+H\right)  \Delta
\gamma-\eta\Delta\dot{\gamma}=0 \label{detgagamma conocido gammadot0}%
\end{align}
so, using a backward Euler evaluation of the normal $~^{tr}f-~^{t}f=2\mu
c\left(  ^{tr}\boldsymbol{\hat{n}}:\Delta\boldsymbol{\varepsilon}\right)
$---this can be seen as the inverse of the relaxation case%
\begin{align}
\Delta\gamma &  =\frac{~\left(  ^{tr}f-~^{t}f\right)  -\eta\Delta\dot{\gamma}%
}{2\mu c^{2}+H}=\dfrac{2\mu c\left(  ^{tr}\boldsymbol{\hat{n}}:\Delta
\boldsymbol{\varepsilon}\right)  }{2\mu c^{2}+H}-\frac{\eta\Delta\dot{\gamma}%
}{2\mu c^{2}+H}\label{detgagamma conocido gammadot}\\
&  =\Delta\gamma_{\infty}-\hat{\tau}\Delta\dot{\gamma}%
\end{align}
Using Eq. (\ref{gammadot2}) into Eq. (\ref{detgagamma conocido gammadot})%
\begin{align}
\Delta\gamma &  =\frac{^{tr}f-~^{t}f}{2\mu c^{2}+H}-\frac{\eta}{2\mu c^{2}%
+H}\left[  \dfrac{2\mu c\left(  ^{tr}\boldsymbol{\hat{n}}:\Delta
\boldsymbol{\varepsilon}/\Delta t\right)  }{2\mu c^{2}+H}-~^{t}\dot{\gamma
}\right]  \left[  1-\exp\left(  -\dfrac{2\mu c^{2}+H}{\eta}\Delta t\right)
\right] \nonumber\\
&  =\underset{%
\begin{array}
[c]{c}%
\text{inviscid}%
\end{array}
}{\underbrace{\Delta\gamma_{\infty}}}\underset{%
\begin{array}
[c]{c}%
\text{viscous}%
\end{array}
}{\underbrace{-\hat{\tau}~^{tr}\dot{\gamma}_{neq}\left[  1-\exp\left(
-\frac{\Delta t}{\hat{\tau}}\right)  \right]  }} \label{dgamman+1}%
\end{align}
\emph{so we recover Eq}. \emph{(\ref{delta gamma}), }but now from $^{t+\Delta
t}f=0$ instead of from integrating directly $\dot\gamma$.

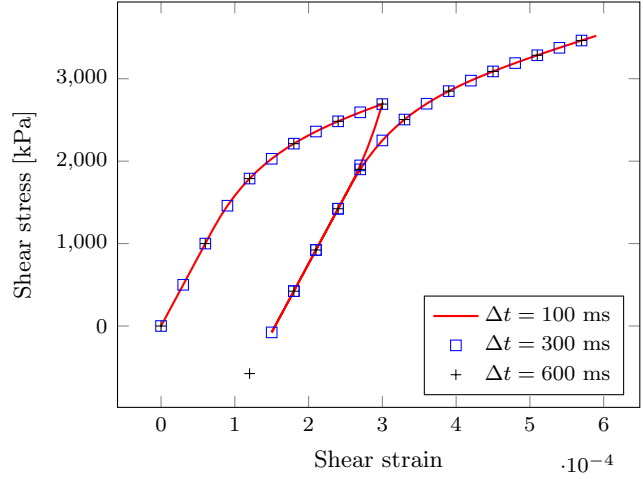
\begin{figure}[h]
\centering
\begin{tikzpicture}
\pgfplotsset{every axis legend/.append style={at={(0.5,1.03)},anchor=south east},}
\begin{axis}[legend style ={cells={anchor=west},legend pos=south east}, ylabel={Shear stress [kPa]},
xlabel={Shear strain},mark repeat=1]
\addplot[thick,color=red]  table [x expr=\thisrowno{1}, y expr=\thisrowno{2}]{results_pure_shear_dt_1.0e-5_loading-unloading.txt};
\addplot[only marks, mark=square, blue]  table [x expr=\thisrowno{1}, y expr=\thisrowno{2}]{results_pure_shear_dt_3.0e-5_loading-unloading.txt};
\addplot[only marks, mark=+, color=black]  table [x expr=\thisrowno{1}, y expr=\thisrowno{2}]{results_pure_shear_dt_6.0e-5_loading-unloading.txt};
\legend{$\Delta t = 100$ ms, $\Delta t=300$ ms, $\Delta t=600$ ms}
\end{axis}
\end{tikzpicture}
\caption{Loading-unloading-reloading with different time increments using the
proposed algorithm. Note that viscoplastic solutions are coincident at
specific computational points (step ends) regardless of the time-increment
employed and of the lack of coincidence of step ends when crossing the yield
surface or when loading is reversed. }%
\label{fig.exact_solution}%
\end{figure}

Summarizing, the solution for the linear viscoplastic problem is given by the
system of equations given by Eqs. (\ref{gammadot2}) and (\ref{delta gamma}).
Note that this solution recovers automatically those when $\eta=0$ (inviscid
plasticity) and when $\kappa=H=0$ (viscoelasticity). Remarkably, the solutions
of $\Delta\gamma$ and $\Delta\dot{\gamma}$ in Eqs. (\ref{gammadot2}) and
(\ref{delta gamma}) are the \emph{exact} solutions that fulfill,
\textit{during all the step,} the thermodynamic consistency given by $f=0$
from $t$ to $t+\Delta t$, with the requirements that: (1) elasticity moduli
$\mu$, hardening $H$ and viscosity $\eta$ are constant, and (2) the rate $2\mu
c\left(  \boldsymbol{\hat{n}}:\boldsymbol{\dot{\varepsilon}}\right)  $ is
constant during the step. In proportional loading in linear viscoplasticity,
these are fulfilled. Figure \ref{fig.exact_solution} shows that the same
solution is obtained for different time step increments. In other cases
(nonlinear viscoelasticity or multiaxial non-proportional loading), the
present solution is only an approximation, and backward-Euler evaluations are
employed to recover the inviscid solution for $\Delta t\rightarrow\infty$ or
$\eta=0$.

Once the values of $^{t+\Delta t}\gamma$ and $^{t+\Delta t}\dot{\gamma}$ are
known, the elastic strain is computed from a backward-Euler scheme as%
\begin{equation}
^{t+\Delta t}\boldsymbol{\varepsilon}_{e}=\underset{%
\begin{array}
[c]{c}%
~^{t}\boldsymbol{\varepsilon}_{e}+\Delta^{tr}\boldsymbol{\varepsilon}_{e}%
\end{array}
}{\underbrace{~^{t}\boldsymbol{\varepsilon}_{e}+\Delta\boldsymbol{\varepsilon
}}}\underset{_{%
\begin{array}
[c]{c}%
+\Delta^{ct}\boldsymbol{\varepsilon}_{e}%
\end{array}
}}{\underbrace{-\Delta\gamma c~^{t+\Delta t}\boldsymbol{\hat{n}}}}
\label{59eq}%
\end{equation}
and obviously from the hyperelastic relation with $\mathbb{C}_{e}=d^{2}%
\Psi/d\boldsymbol{\varepsilon}_{e}\otimes d\boldsymbol{\varepsilon}_{e}$%
\begin{equation}
^{t+\Delta t}\boldsymbol{\sigma}=\mathbb{C}_{e}:~^{t+\Delta t}%
\boldsymbol{\varepsilon}_{e}=Ktr\left(  ^{t+\Delta t}\boldsymbol{\varepsilon
}_{e}\right)  \boldsymbol{I}+2\mu~^{t+\Delta t}\boldsymbol{\varepsilon}%
_{e}^{d}%
\end{equation}
Consequently, the consistent tangent modulus tensor during the step fully
viscoplastic can be determined as
\begin{equation}
^{t+\Delta t}\mathbb{C}=\dfrac{d~^{t+\Delta t}\boldsymbol{\sigma}%
}{d~^{t+\Delta t}\boldsymbol{\varepsilon}}=~^{t+\Delta t}\mathbb{C}%
^{v}+\mathbb{~}^{t+\Delta t}\mathbb{C}^{d}=K\boldsymbol{I}\otimes
\boldsymbol{I}+\dfrac{d~^{t+\Delta t}\boldsymbol{\sigma}^{d}}{d~^{t+\Delta
t}\boldsymbol{\varepsilon}} \label{eqdsde}%
\end{equation}
From $^{t+\Delta t}\boldsymbol{\varepsilon}_{e}^{d}=~^{tr}%
\boldsymbol{\varepsilon}_{e}^{d}-c\Delta\gamma~^{t+\Delta t}\boldsymbol{\hat
{n}}$,
\begin{equation}
\dfrac{d~^{t+\Delta t}\boldsymbol{\varepsilon}_{e}^{d}}{d~^{t+\Delta
t}\boldsymbol{\varepsilon}}=\mathbb{P}^{s}-c\Delta\gamma\dfrac{d~^{t+\Delta
t}\boldsymbol{\hat{n}}}{d~^{t+\Delta t}\boldsymbol{\varepsilon}}-c~^{t+\Delta
t}\boldsymbol{\hat{n}}\otimes\frac{d\Delta\gamma}{d~^{t+\Delta t}%
\boldsymbol{\varepsilon}}%
\end{equation}
with---recall that $~^{t+\Delta t}\boldsymbol{\hat{n}\equiv}$ $~^{tr}%
\boldsymbol{\hat{n}=~}^{tr}\boldsymbol{\sigma}^{d}/||^{tr}\boldsymbol{\sigma
}^{d}||$%
\begin{equation}
\mathbb{N}:=\dfrac{d~^{t+\Delta t}\boldsymbol{\hat{n}}}{d~^{t+\Delta
t}\boldsymbol{\varepsilon}}=\dfrac{d~^{tr}\boldsymbol{\hat{n}}}{d~^{t+\Delta
t}\boldsymbol{\varepsilon}}=\dfrac{2\mu}{||^{tr}\boldsymbol{\sigma}^{d}%
||}\left[  {\ \mathbb{P}^{s}-~}^{tr}{\boldsymbol{\hat{n}}\otimes~^{t+\Delta
t}\boldsymbol{\hat{n}}}\right]
\end{equation}
Using the conditions $^{t+\Delta t}f=0$ and $^{t+\Delta t}\dot{\gamma}$ from
Eq. (\ref{gammadotn+1}), after some straightforward algebra, we arrive at
\begin{equation}
\,^{t+\Delta t}\mathbb{C}^{d}:=\dfrac{d~^{t+\Delta t}\boldsymbol{\sigma}^{d}%
}{d~^{t+\Delta t}\boldsymbol{\varepsilon}}=2\mu\left[  1-\dfrac{2\mu}%
{2\mu+H/c^{2}}\left(  1-\dfrac{\hat{\tau}}{\Delta t}~\xi(\Delta t)\right)
\right]  \mathbb{K}^{-1}:\mathbb{P}^{s}%
\end{equation}
in which $\mathbb{K}$ is
\begin{equation}
\mathbb{K}=\mathbb{I}^{S}+2\mu c\hat{\tau}~^{t}\dot{\gamma}~\xi(\Delta
t)~\mathbb{N}%
\end{equation}

Of course, for this linear case, the tangent developed below for the nonlinear
case may be equally used. Note that $\mathbb{C}^{d}$ is also bounded by the
deviatoric elastic tangent modulus tensor $\mathbb{C}_{e}^{d}$ and by the
deviatoric consistent inviscid elastoplastic tangent modulus tensor
$\mathbb{C}_{ep}^{d}$ as shown in the continuum theory. In fact,
\begin{equation}
\mathbb{C}^{d}=%
\begin{cases}
2\mu\mathbb{P}^{s}\equiv\mathbb{C}_{e}^{d} & \quad\text{for}\quad\Delta
t\rightarrow0,\eta\rightarrow\infty\\
\dfrac{2\mu H/c^{2}}{2\mu+H/c^{2}}\mathbb{P}^{s}\equiv\mathbb{C}_{ep}^{d} &
\quad\text{for}\quad\Delta t\rightarrow\infty,\eta\rightarrow0
\end{cases}
\end{equation}

\subsection{Crossing the elastic domain limit}

An important algorithmic issue is when a step is crossing the limit of the
elastic domain, i.e. when it is initially elastic but ends being viscoplastic,
or vice-versa (unloading). Assuming that time step $t$ has no instantaneous
viscoplastic flow (i.e. the previous step ended elastic), then $^{t}%
\dot{\gamma}=0$. If $^{t}\dot{\gamma}=0$, the step will be elastic unless
$^{tr}f\equiv~^{tr}f_{p}>0$, because the condition $^{t+\Delta t}f>0$ is not
possible. However, if $^{t}f_{p}<0$, some part of the step is\ still elastic.
In contrast to perfect plasticity, since speed affects the solution, the step
must be partitioned to identify which part is dissipative if we want the exact
solution for the linear proportional loading case. Indeed, the predictions in
Fig. \ref{fig.exact_solution} have been obtained using these partitions. Then
consider the following partition%
\begin{equation}
\left\{
\begin{array}
[c]{l}%
\Delta t\text{ }=\text{ total step }= \Delta t^{c} + \Delta t^{*}\\
\Delta t^{c}=\text{ conservative part of the step}\\
\Delta t^{\ast}=\text{dissipative part of the step}%
\end{array}
\right.
\end{equation}
and apply the nomenclature to all variables, i.e. $\left(  \bullet\right)
^{c}$ is for the conservative part of the step, and $\left(  \bullet\right)
^{\ast}$ is for the dissipative part of the step. Recall that $\Delta
\boldsymbol{\varepsilon}/\Delta t$ is constant during all the step, so we can
write%
\begin{equation}
\frac{\Delta\gamma_{\infty}}{\Delta t}=\dfrac{2\mu c\left(  ^{tr}%
\boldsymbol{\hat{n}}:\Delta\boldsymbol{\varepsilon}/\Delta t\right)  }{2\mu
c^{2}+H}=\dfrac{2\mu c\left(  ^{tr}\boldsymbol{\hat{n}}:\Delta
\boldsymbol{\varepsilon}^{\ast}/\Delta t^{\ast}\right)  }{2\mu c^{2}+H}%
=\frac{\Delta\gamma_{\infty}^{\ast}}{\Delta t^{\ast}}=\dfrac{2\mu c\left(
^{tr}\boldsymbol{\hat{n}}:\Delta\boldsymbol{\varepsilon}^{c}/\Delta
t^{c}\right)  }{2\mu c^{2}+H}=\frac{\Delta\gamma_{\infty}^{c}}{\Delta t^{c}}%
\end{equation}
i.e. $\dot{\gamma}_{\infty}=\dot{\gamma}_{\infty}^{c}=\dot{\gamma}_{\infty
}^{\ast}$. The first part of the step is given by $\Delta t^{c}$ such that
$^{t+\Delta t^{c}}f=0$, but $^{\tau}f<0$, $\forall\tau\in\left(  t,t+\Delta
t^{c}\right)  $. The second sub-step, with $\Delta t^{\ast}$ gives%
\begin{equation}
\Delta\gamma^{\ast}\equiv\Delta\gamma=\dot{\gamma}_{\infty}\Delta t^{\ast
}-\dot{\gamma}_{\infty}\hat{\tau}~\xi(\Delta t^{\ast})
\end{equation}
and since $^{t}\dot{\gamma}=~^{t+\Delta t^{c}}\dot{\gamma}=0$, Eq.
(\ref{gammadot2}) gives%
\begin{equation}
^{t+\Delta t}\dot{\gamma}=\dot{\gamma}_{\infty}~\,\xi(\Delta t^{\ast})
\end{equation}
where $\Delta t^{\ast}$ is unknown, but can be obtained from $^{t+\Delta
t}f=0$ with $~^{tr}\boldsymbol{\sigma}=d\Psi(^{tr}\boldsymbol{\varepsilon}%
_{e})/d^{tr}\boldsymbol{\varepsilon}_{e}$ i.e.%
\begin{equation}
^{t+\Delta t}f\equiv\;\overset{%
\begin{array}
[c]{c}%
^{tr}f
\end{array}
}{\overbrace{\left[  c~^{tr}\boldsymbol{\sigma}:~^{tr}\boldsymbol{\hat{n}%
}-\kappa_{0}-H~^{t}\gamma\right]  }}-\left(  2\mu c^{2}+H\right)
{\Delta\gamma}^{\ast}-\eta~^{t+\Delta t}\dot{\gamma}=0
\end{equation}
so%
\begin{equation}
^{t+\Delta t}f\equiv~^{tr}f-\left(  2\mu c^{2}+H\right)  \left[  \dot{\gamma
}_{\infty}\Delta t^{\ast}-\dot{\gamma}_{\infty}\hat{\tau}\Delta t^{\ast}%
~\xi(\Delta t^{\ast})\right]  -\eta\dot{\gamma}_{\infty}~\xi(\Delta t^{\ast
})=0 \label{eq.dissptime}%
\end{equation}
Note that for the case $\hat{\tau}=0$ we have $~\xi(\Delta t^{\ast})=\left[
1-\exp\left(  -\Delta t^{\ast}/\hat{\tau}\right)  \right]  =0$ and%
\begin{equation}
^{t+\Delta t}f\equiv~^{tr}f-\left(  2\mu c^{2}+H\right)  \dot{\gamma}_{\infty
}\Delta t^{\ast}=0\;\Rightarrow\;\Delta t^{\ast}=\frac{~^{tr}f}{\left(  2\mu
c^{2}+H\right)  \dot{\gamma}_{\infty}}=\frac{~^{tr}f}{2\mu c\left(
^{tr}\boldsymbol{\hat{n}}:\Delta\boldsymbol{\varepsilon}/\Delta t\right)  }
\label{eqdissptime2}%
\end{equation}
which gives the correct partition of the step in the computation of the
dissipative part and conservative parts in inviscid elastoplasticity. In the
more general case, Eq. (\ref{eq.dissptime}) needs to be solved for
numerically, e.g. using a Newton-Raphson scheme, with tangent%
\begin{equation}
\dfrac{d^{t+\Delta t}f}{d\Delta t^{\ast}}=-2\mu c\left(  ^{tr}\boldsymbol{\hat
{n}}:\Delta\boldsymbol{\varepsilon}/\Delta t\right)  \{1-\hat{\tau}[\xi(\Delta
t^{\ast})+\Delta t^{\ast}\xi^{\prime\ast})]-\eta\xi^{\prime\ast})\}
\end{equation}
where $\xi(\Delta t^{\ast})=1-\exp(-\Delta t^{\ast}/\hat{\tau})$ and
$\xi^{\prime\ast}=-({1}/{\hat{\tau}})\exp(-\Delta t^{\ast}/\hat{\tau})$. Note
that the case $\eta=\hat{\tau}=0$ is automatically recovered by the first
iteration in a Newton-Raphson method, e.g. it results in Eq.
(\ref{eqdissptime2}) if we depart from a first guess $~^{t+\Delta t}%
f^{[0]}=~^{tr}f$. Note also that if $\Delta t^{\ast}=0$, then $\xi(\Delta
t^{\ast}=0)=0$ and $\xi^{\prime\ast}(\Delta t^{\ast}=0)=1/\hat{\tau}$, so
\begin{equation}
\dfrac{d^{t+\Delta t}f}{d\Delta t^{\ast}}=-2\mu c\left(  ^{tr}\boldsymbol{\hat
{n}}:\Delta\boldsymbol{\varepsilon}/\Delta t\right)  \{1-2\mu c^{2}-H\}
\end{equation}
which is the inviscid solution, because in such case $\dot{\gamma}=0$.

\subsection{Unloading case}

In contrast to inviscid plasticity, $\,^{tr}f_{p}<0$ does not imply
that the step ends up being elastic. As aforementioned, instead of the
classical Kuhn-Tucker condition, the unloading case is detected by the
computation of a resulting $^{t+\Delta t}\dot{\gamma}<0$ from a usual
viscoplastic step, namely%
\begin{equation}
^{t+\Delta t}\dot{\gamma}=\underbrace{\dot{\gamma}_{\infty}}_{<0}%
+\underbrace{\left(  ^{t}\dot{\gamma}-{\dot{\gamma}_{\infty}}\right) }_{>0}
\exp\left(  -\frac{\Delta t}{\hat{\tau}}\right)  <0
\end{equation}
Note that we may have $\dot\gamma_{\infty}<0$ but a final $^{t+\Delta t}%
\dot{\gamma}<0$ is not a possible solution. Note that after reversing loading,
$\,^{t}\dot\gamma_{neq}=\dot\gamma_{\infty}-\,^{t}\dot\gamma\lneqq0$, see Fig.
\ref{dibujos}f, where we seek to find the instant at $C^{\prime}$. Then, for
an accurate solution, we need to divide the step in a first sub-step $\Delta
t^{\ast}$ in which dissipation takes place and a second sub-step $\Delta
t^{c}$ in which no dissipation takes place. The size of the first sub-step is
computed precisely from that condition using, for example, the residual%
\begin{equation}
r_{t}:=\Delta t^{\ast}-\hat{\tau}_{0}\log\left(  \frac{~^{\ast}\dot{\gamma
}_{\infty}-~^{t}\dot{\gamma}}{~^{\ast}\dot{\gamma}_{\infty}}\right)  =0\text{
\ with }\Delta t^{\ast}\in\left(  0,\Delta t\right)  \label{rt}%
\end{equation}
where%
\begin{equation}
~^{\ast}{\dot{\gamma}_{\infty}=}\dfrac{2\mu c\left(  ^{\ast}\boldsymbol{\hat
{n}}:\Delta\boldsymbol{\varepsilon}/\Delta t\right)  }{2\mu c^{2}+H}<0
\end{equation}
and $^{\ast}\boldsymbol{\hat{n}}\neq~^{tr}\boldsymbol{\hat{n}}$ is the normal
when crossing the plastic yield surface $f_{p}$, i.e. when $^{t+\Delta t}%
\dot{\gamma}\equiv~^{\ast}\dot{\gamma}=0$ (end of the viscoplastic substep and
start of the elastic unloading)
\begin{equation}
^{\ast}\boldsymbol{\hat{n}}=\frac{^{\ast}\boldsymbol{\sigma}^{d}\left(  \Delta
t^{\ast}\right)  }{\left\Vert ^{\ast}\boldsymbol{\sigma}^{d}\left(  \Delta
t^{\ast}\right)  \right\Vert }\text{ with }^{\ast}\boldsymbol{\sigma}%
^{d}\left(  \Delta t^{\ast}\right)  =~^{t}\boldsymbol{\sigma}^{d}+\frac{\Delta
t^{\ast}}{\Delta t}2\mu\Delta\boldsymbol{\varepsilon}^{d}%
\end{equation}
The scalar nonlinear Equation (\ref{rt}) is solved iteratively using any
suitable method, e.g. a Newton-Raphson method, for which the tangent is%
\begin{equation}
\frac{dr_{t}\left(  \Delta t^{\ast}\right)  }{d\Delta t^{\ast}}=1-\frac
{~^{t}\dot{\gamma}~\hat{\tau}_{0}}{~^{\ast}\dot{\gamma}_{\infty}\left(
~^{\ast}\dot{\gamma}_{\infty}-~^{t}\dot{\gamma}\right)  }\frac{d~^{\ast}%
{\dot{\gamma}_{\infty}}}{d\Delta t^{\ast}}%
\end{equation}
with%
\begin{equation}
\frac{d~^{\ast}{\dot{\gamma}_{\infty}}}{d\Delta t^{\ast}}=\frac{1}{\Delta
t}\dfrac{2\mu c}{2\mu c^{2}+H}\Delta\boldsymbol{\varepsilon}:\frac{d~^{\ast
}\boldsymbol{\hat{n}}}{d\Delta t^{\ast}}%
\end{equation}
and%
\begin{equation}
\frac{d~^{\ast}\boldsymbol{\hat{n}}}{d\Delta t^{\ast}}=\frac{2\mu}{\Delta
t}\frac{1}{\left\Vert ^{\ast}\boldsymbol{\sigma}^{d}\left(  \Delta t^{\ast
}\right)  \right\Vert }(\mathbb{P}^{s}-~^{\ast}\boldsymbol{\hat{n}\otimes
}^{\ast}\boldsymbol{\hat{n}}):\Delta\boldsymbol{\varepsilon}^{d}%
\end{equation}
The iterations are%
\begin{equation}
\left[  \Delta t^{\ast}\right]  ^{\left(  j+1\right)  }=\left[  \Delta
t^{\ast}\right]  ^{\left(  j\right)  }-\left[  \frac{dr^{\ast}\left(  \Delta
t^{\ast}\right)  }{d\Delta t^{\ast}}\right]  ^{\left(  j\right)  -1}%
r^{\ast\left(  j\right)  }%
\end{equation}
and the first guess may be obtained using $(\Delta t^{*})^{0}=0$ and
\begin{equation}
~^{\ast}{\dot{\gamma}_{\infty}^{\left(  0\right)  }:=}\dfrac{2\mu c\left(
^{t}\boldsymbol{\hat{n}}:\Delta\boldsymbol{\varepsilon}/\Delta t\right)
}{2\mu c^{2}+H}%
\end{equation}
Thereafter%
\begin{equation}
\Delta\gamma^{\ast}\equiv\Delta\gamma=\Delta^{\ast}\gamma_{\infty}+\hat{\tau
}\left(  ^{t}\dot{\gamma}-~^{\ast}\dot{\gamma}_{\infty}\right)  \left[
1-\exp\left(  -\frac{\Delta t^{\ast}}{\hat{\tau}}\right)  \right]
\end{equation}
Then, the remaining part of the sub-step $\Delta t^{c}=\Delta t-\Delta
t^{\ast}$ is elastic, with a deviatoric strain increment of%
\begin{equation}
\Delta\boldsymbol{\varepsilon}^{d\,c}=\frac{\Delta t^{c}}{\Delta t}%
\Delta\boldsymbol{\varepsilon}^{d}%
\end{equation}
However, note that the elastic strains are computed from the trial ones
directly as%
\begin{equation}
^{t+\Delta t}\boldsymbol{\varepsilon}_{e}=\underset{%
\begin{array}
[c]{c}%
^{tr}\boldsymbol{\varepsilon}_{e}%
\end{array}
}{\underbrace{~^{t}\boldsymbol{\varepsilon}_{e}+\Delta\boldsymbol{\varepsilon
}}}-\Delta\gamma^{\ast}~^{\ast}\boldsymbol{\hat{n}}%
\end{equation}

\subsection{Partitioned tangents}

In the cases when the steps include sub-steps, we need a special, partitioned
computation of the tangent. The partition of the step is%
\begin{equation}
\Delta\boldsymbol{\varepsilon}=\Delta\boldsymbol{\varepsilon}^{c}%
+\Delta\boldsymbol{\varepsilon}^{\ast}=\frac{\Delta\boldsymbol{\varepsilon}%
}{\Delta t}\Delta t^{c}+\frac{\Delta\boldsymbol{\varepsilon}}{\Delta t}\Delta
t^{\ast} \label{time_relations}%
\end{equation}
Then, if $\mathbb{C}_{vp}(\Delta t^{*})$ is the viscoplastic tangent for a
step of size $\Delta t^{*}$%
\begin{align}
\left.  ^{t+\Delta t}\mathbb{C}=\frac{d^{t+\Delta t}\boldsymbol{\sigma}%
}{d^{t+\Delta t}\boldsymbol{\varepsilon}}=\frac{d\Delta\boldsymbol{\sigma}%
}{d\Delta\boldsymbol{\varepsilon}}\right.   &  =\frac{d\Delta
\boldsymbol{\sigma}}{d\Delta\boldsymbol{\varepsilon}^{c}}:\frac{d\Delta
\boldsymbol{\varepsilon}^{c}}{d\Delta\boldsymbol{\varepsilon}}+\frac
{d\Delta\boldsymbol{\sigma}}{d\Delta\boldsymbol{\varepsilon}^{\ast}}%
:\frac{d\Delta\boldsymbol{\varepsilon}^{\ast}}{d\Delta\boldsymbol{\varepsilon
}}\\
&  =\mathbb{C}_{e}:\frac{d\Delta\boldsymbol{\varepsilon}^{c}}{d\Delta
\boldsymbol{\varepsilon}}+\mathbb{C}_{vp}(\Delta t^{\ast}):\frac
{d\Delta\boldsymbol{\varepsilon}^{\ast}}{d\Delta\boldsymbol{\varepsilon}}%
\end{align}
where, using Eq. (\ref{time_relations})%
\begin{equation}
\frac{d\Delta\boldsymbol{\varepsilon}^{c}}{d\Delta\boldsymbol{\varepsilon}%
}=\left.  \frac{\partial\Delta\boldsymbol{\varepsilon}^{c}}{\partial
\Delta\boldsymbol{\varepsilon}}\right\vert _{\Delta t^{c}=\text{const}%
}+\left.  \frac{\partial\Delta\boldsymbol{\varepsilon}^{c}}{\partial\Delta
t^{c}}\right\vert _{\Delta\boldsymbol{\varepsilon=}\text{const}}\otimes
\frac{\partial\Delta t^{c}}{\partial\Delta\boldsymbol{\varepsilon}}%
=\frac{\Delta t^{c}}{\Delta t}\mathbb{I}^{S}+\frac{\Delta
\boldsymbol{\varepsilon}}{\Delta t}\boldsymbol{\otimes}\frac{\partial\Delta
t^{c}}{\partial\Delta\boldsymbol{\varepsilon}}%
\end{equation}
and
\begin{equation}
\frac{d\Delta\boldsymbol{\varepsilon}^{\ast}}{d\Delta\boldsymbol{\varepsilon}%
}=\left.  \frac{\partial\Delta\boldsymbol{\varepsilon}^{\ast}}{\partial
\Delta\boldsymbol{\varepsilon}}\right\vert _{\Delta t^{\ast}=\text{const}%
}+\left.  \frac{\partial\Delta\boldsymbol{\varepsilon}^{\ast}}{\partial\Delta
t^{\ast}}\right\vert _{\Delta\boldsymbol{\varepsilon=}\text{const}}%
\otimes\frac{\partial\Delta t^{\ast}}{\partial\Delta\boldsymbol{\varepsilon}%
}=\frac{\Delta t^{\ast}}{\Delta t}\mathbb{I}^{S}+\frac{\Delta
\boldsymbol{\varepsilon}}{\Delta t}\boldsymbol{\otimes}\frac{\partial\Delta
t^{\ast}}{\partial\Delta\boldsymbol{\varepsilon}}%
\end{equation}
where $\partial\Delta t^{c}/\partial\Delta\boldsymbol{\varepsilon}$ and
$\partial\Delta t^{\ast}/\partial\Delta\boldsymbol{\varepsilon}$ are obtained
from the respective conditions of $f=0$ and $\dot{\gamma}=0$ (depending on the
condition governing the step partitioning), and once one condition is
obtained, the other one is given by the complementarity of the other substep
step; for example%
\begin{equation}
\Delta t=\Delta t^{c}+\Delta t^{\ast}\Rightarrow\frac{\partial\Delta
t}{\partial\Delta\boldsymbol{\varepsilon}}=\frac{\partial\Delta t^{c}%
}{\partial\Delta\boldsymbol{\varepsilon}}+\frac{\partial\Delta t^{\ast}%
}{\partial\Delta\boldsymbol{\varepsilon}}=\boldsymbol{0}%
\end{equation}
because $\Delta t$ is constant, independent of $\Delta\boldsymbol{\varepsilon
}$, so%
\begin{equation}
\frac{\partial\Delta t^{\ast}}{\partial\Delta\boldsymbol{\varepsilon}}%
=-\frac{\partial\Delta t^{c}}{\partial\Delta\boldsymbol{\varepsilon}}%
\end{equation}

Here, we develop $\partial\Delta t^{\ast}/\partial\Delta
\boldsymbol{\varepsilon} $ for two cases: one starts initially elastic and
ends being viscoplastic (see Sec. 3.2), other starts initially viscoplastic
but ends being elastic (see Sec. 3.3).

\subsubsection{ First case: crossing the elastic domain to the viscoplastic
domain}

In order to determine the $\Delta t^{\ast}$, we need to solve the nonlinear
Eq. \eqref{eq.dissptime}, which analytical closed-form solution is not easy to
obtain, so a numerical one through the Newton-Raphson method is obtained. Once
the solution is converged, Eq. \eqref{eq.dissptime} is fulfilled and
$\partial\Delta t^{\ast}/\partial\Delta\boldsymbol{\varepsilon}$ can be
obtained by deriving Eq. \eqref{eq.dissptime} respect to $\Delta
\boldsymbol{\varepsilon}$, which after some straightforward math gives
\begin{equation}
\dfrac{\partial\Delta t^{\ast}}{\partial\Delta\boldsymbol{\varepsilon}}%
=\dfrac{1}{\zeta}\dfrac{\partial~^{tr}f}{\partial\Delta\boldsymbol{\varepsilon
}}%
\end{equation}
with%
\begin{equation}
\zeta=(2\mu c^{2}+H)\dot{\gamma}_{\infty}\left\{  1-\hat{\tau}\left[
\xi(\Delta t^{\ast})+\frac{\Delta t^{\ast}}{\hat{\tau}}\exp\left(
-\frac{\Delta t^{\ast}}{\hat{\tau}}\right)  \right]  \right\}  +\frac{\eta
\dot{\gamma}_{\infty}}{\hat{\tau}}\exp\left(  -\frac{\Delta t^{\ast}}%
{\hat{\tau}}\right)
\end{equation}
and
\begin{equation}
\dfrac{\partial~^{tr}f}{\partial\Delta\boldsymbol{\varepsilon}}=2\mu c\left(
\Delta\boldsymbol{\varepsilon}:\mathbb{P}_{n}:\mathbb{C}^{e}+~^{tr}%
\boldsymbol{\hat{n}}\right)
\end{equation}

\subsubsection{Second case: crossing the viscoplastic domain to the elastic
domain}

From the established relation in Eq. \eqref{rt},
\begin{equation}
\frac{dr_{t}\left(  \Delta\boldsymbol{\varepsilon},\Delta t^{\ast}\left(
\Delta\boldsymbol{\varepsilon}\right)  \right)  }{d\Delta
\boldsymbol{\varepsilon}}=\left.  \frac{\partial r_{t}\left(  \Delta t^{\ast
}\right)  }{\partial\Delta t^{\ast}}\right\vert _{\Delta\varepsilon
=\text{const}}\dfrac{\partial\Delta t^{\ast}}{\partial\Delta
\boldsymbol{\varepsilon}}+\left.  \frac{\partial r_{t}\left(  \Delta t^{\ast
}\right)  }{\partial\Delta\boldsymbol{\varepsilon}}\right\vert _{\Delta
t^{\ast}=\text{const}}=\mathbf{0}%
\end{equation}
so%
\begin{equation}
\dfrac{\partial\Delta t^{\ast}}{\partial\Delta\boldsymbol{\varepsilon}%
}=-\left[  \left.  \frac{\partial r_{t}\left(  \Delta t^{\ast}\right)
}{\partial\Delta t^{\ast}}\right\vert _{\Delta\varepsilon=\text{const}%
}\right]  ^{-1}\left.  \frac{\partial r_{t}\left(  \Delta t^{\ast}\right)
}{\partial\Delta\boldsymbol{\varepsilon}}\right\vert _{\Delta t^{\ast
}=\text{const}}%
\end{equation}
where%
\begin{equation}
\left.  \frac{\partial r_{t}\left(  \Delta t^{\ast}\right)  }{\partial
\Delta\boldsymbol{\varepsilon}}\right\vert _{\Delta t^{\ast}=\text{const}%
}=-\frac{~^{t}\dot{\gamma}~\hat{\tau}_{0}}{~^{\ast}\dot{\gamma}_{\infty
}\left(  ~^{\ast}\dot{\gamma}_{\infty}-~^{t}\dot{\gamma}\right)  }\left.
\frac{\partial~^{\ast}{\dot{\gamma}_{\infty}}}{\partial\Delta
\boldsymbol{\varepsilon}}\right\vert _{\Delta t^{\ast}=\text{const}}%
\end{equation}
with%
\begin{equation}
\left.  \frac{\partial~^{\ast}{\dot{\gamma}_{\infty}}}{\partial\Delta
\boldsymbol{\varepsilon}}\right\vert _{\Delta t^{\ast}=\text{const}}=\frac
{1}{\Delta t}\dfrac{2\mu c}{2\mu c^{2}+H}~^{\ast}\boldsymbol{\hat{n}+}\frac
{1}{\Delta t}\dfrac{2\mu c}{2\mu c^{2}+H}\Delta\boldsymbol{\varepsilon
}:\left.  \frac{\partial~^{\ast}\boldsymbol{\hat{n}}}{\partial\Delta
\boldsymbol{\varepsilon}}\right\vert _{\Delta t^{\ast}=\text{const}}%
\end{equation}
and%
\begin{equation}
\left.  \frac{\partial~^{\ast}\boldsymbol{\hat{n}}}{\partial\Delta
\boldsymbol{\varepsilon}}\right\vert _{\Delta t^{\ast}=\text{const}}%
=\frac{2\mu\Delta t^{\ast}/\Delta t}{\left\Vert ^{\ast}\boldsymbol{\sigma}%
^{d}\left(  \Delta t^{\ast}\right)  \right\Vert }(\mathbb{P}^{s}-~^{\ast
}\boldsymbol{\hat{n}}\otimes~^{\ast}\boldsymbol{\hat{n}}) \label{parcial_e1}%
\end{equation}

\section{Comparison with classical models}

Frequently, different interpretations of the rheological model of Fig.
\ref{fig.binghammodel} are considered as different models or formulations in
the literature, even though in practice they may correspond to the same
physics. However, equations are typically arranged in different ways so they
become more convenient for specific purposes, allowing different
interpretations and specially different algorithmic schemes, which are of most
importance in finite element analysis.

\subsection{Perzyna formulation}

The model from Perzyna \cite{Perzyna1966}, with different variations, is
probably the best known model in computational viscoplasticity. The main asset
of the model is the simplicity, because it does not require the fulfillment of
the so-called consistency condition. The main handicap is the bad conditioning
obtained as $\eta\rightarrow0$, because the model is given by simply stating
the rate as $\dot{\gamma}=\left\langle f_{p}\right\rangle /\eta$, where
$f_{p}$ is the plasticity yield function (i.e. $f$ for $\eta=0$) and
$\langle\bullet\rangle$ is the Macaulay bracket. Hence, the inviscid solution
cannot be recovered. A possible time integration algorithm may be simply
obtained by the formulae $\,^{t+\Delta t}\dot\gamma=\langle^{tr}f_{p}%
\rangle/\eta$ and $\Delta\gamma=\,^{t}\gamma+\,^{t+\Delta t}\dot\gamma\Delta
t$. Perzyna's model is also frequently written using a dimensionless viscosity
parameter $\bar{\eta}$, an exponent $N\equiv1/\epsilon\geq1$, and a
nondimensional inviscid yield function $f_{p}/\bar{\kappa}$, with $\bar
{\kappa}$ being the nondimensionalization factor. This is the so-called power
model%
\begin{equation}
\dot{\gamma}=\frac{\left\langle \phi\left(  f_{p}\right)  \right\rangle }%
{\bar{\eta}}\text{ with }\phi\left(  f_{p}\right)  :=\left(  \frac{f_{p}}%
{\bar{\kappa}}\right)  ^{N} \label{Perzyna}%
\end{equation}
which is also undefined for $\bar{\eta}=0$ (hence the source of numerical
problems in some implementations). For simplicity in the comparison we use the
(constant, initial) value $\bar{\kappa}=~^{0}\kappa$ (this factor is included
only in some formulations). In this case Eq. (\ref{Perzyna}) may be re-written
as
\begin{equation}
f:=f_{p}-~\bar{\kappa}\bar{\eta}^{1/N}\dot{\gamma}^{1/N}=0 \label{eqfPerzyna}%
\end{equation}
i.e. we can write the energy conservation principle as%
\begin{equation}
f:=f_{p}-g\left(  \dot{\gamma}\right)  =0\text{ \ with }g\left(  \dot{\gamma
}\right)  =~\bar{\kappa}\bar{\eta}^{1/N}\dot{\gamma}^{1/N}=\bar{\kappa}%
\bar{\eta}^{\epsilon}\dot{\gamma}^{\epsilon}%
\end{equation}
so we recover our formulation as given in Eq. (\ref{fdefinition}), and where
the instantaneous viscosity modulus of our formulation is%
\begin{equation}
^{t}\eta=g^{\prime}=\frac{d^{t}g}{d^{t}\dot{\gamma}}=\frac{1}{N}~\bar{\kappa
}\bar{\eta}^{1/N}\dot{\gamma}^{\left(  1/N-1\right)  }=\epsilon\bar{\kappa
}\bar{\eta}^{\epsilon}\dot{\gamma}^{\epsilon-1}%
\end{equation}
For the linear case with $N=\epsilon=1$%
\begin{equation}
\dot{\gamma}=\frac{f_{p}}{\eta}\text{ \ so \ }g\left(  \dot{\gamma}\right)
=\eta\dot{\gamma}^{\text{ }}\text{ and \ }g^{\prime}\equiv\eta=~\bar{\kappa
}\bar{\eta} \label{gammadot perzyna lineal}%
\end{equation}
An issue highlighted by Peric \cite{Peric93}, is that when $\epsilon
\rightarrow0$ for which one would assume to recover an inviscid limit, the
stress approaches the limit $2\bar{\kappa}$. This is apparent particularizing
Eq. (\ref{eqfPerzyna}) for this case, which brings $f(\epsilon\rightarrow
0)\rightarrow f_{p}-\bar{\kappa}=0$ instead of $f_{p}=0$. However, for the
also inviscid limit $\bar{\eta}\rightarrow0$ the correct $f_{p}=0$ is
obtained. A different proposal, given in \cite{Peric93,Souza-Neto} (and
therein references) and in \cite{Miehe}, to overcome the inconsistency in the
sensitivity parameter $\epsilon$, is
\begin{equation}
\dot{\gamma}=\frac{\left\langle \bar{\phi}\left(  f_{p}\right)
-1\right\rangle }{\tilde{\eta}}\text{ with }\bar{\phi}\left(  f_{p}\right)
:=\left(  \frac{f_{p}}{\bar{\kappa}}+1\right)  ^{\tilde{N}}%
\end{equation}
where $\bar{\kappa}$ plays again the role of yield stress. In this case,
following the rehological model, we have
\begin{equation}
f:=f_{p}-g(\dot{\gamma})=f_{p}-\underbrace{\bar{\kappa}[(\dot{\gamma}%
\tilde{\eta}+1)^{1/\tilde{N}}-1]}_{%
\begin{array}
[c]{c}%
g(\dot{\gamma})
\end{array}
}=0
\end{equation}
which, note, recovers the inviscid limit for the cases $\dot{\gamma}=0$,
$\tilde{\eta}=0$ and $1/\tilde{N}=0$, hence the preference for this model in
the computational mechanics literature. In the linear case, we have the same
solution as the Perzyna model, i.e. $\tilde{\eta}=\bar{\eta}=\eta/\bar{\kappa
}$ and $\tilde{N}=N=\epsilon=1$.

In summary, the Perzyna-type models are just a particular case of our
formulation, but our algorithmic solution is well conditioned regardless of
the value of the viscosity $\eta$ (or $\bar{\eta}$). Finally, we note that the
common setting in the materials science literature does not normalize the
yield function nor the viscosity parameter, so they have dimensions of stress.

\subsection{Duvaut-Lions formulation}

Another frequently used formulation in viscoplasticity is the Duvaut-Lions
formulation. Motivated on that framework, other models have also been
presented, see e.g. \cite{Peric93}. The algorithmic advantage of the
Duvaut-Lions model respect to the Perzyna formulation is that the inviscid
case is automatically recovered because, in fact, the viscous solution is
computed as a regularization of the inviscid one, which is computed first. The
model is frequently presented as (see e.g. Eq. (2.7.13) in \cite{Simo1998},
adapted herein to our notation; for example the tensorial $\dot{\gamma}$ in
\cite{Simo1998} is $c\dot{\gamma}$ here because our $\dot{\gamma}$ is the
uniaxial equivalent, and $f$ in \cite{Simo1998} is our $f_{p}/c$)%
\begin{equation}
\dot{c\gamma}=\frac{1}{2\mu\bar{\tau}}\boldsymbol{\hat{n}}:\left(
\boldsymbol{\sigma}^{d}-\frac{~^{t}\kappa}{c}~\boldsymbol{\hat{n}}\right)
\text{ \ if }f_{p}>0\text{; \ }\dot{\gamma}=0\text{ otherwise}
\label{Gammadot DL}%
\end{equation}
Recall that $\boldsymbol{\hat{n}}=\boldsymbol{\sigma}^{d}/\left\Vert
\boldsymbol{\sigma}^{d}\right\Vert $ and $\bar{\tau}$ is a relaxation time.
Note that in Eq. (\ref{Gammadot DL}) $\boldsymbol{\sigma}$ is the stress,
which may be outside the inviscid yield surface, and since $\boldsymbol{\sigma
}^{d}$ has the direction $\boldsymbol{\hat{n}}$, and $^{t}\kappa$ is the
inviscid uniaxial yield stress, $^{t}\kappa/c~\boldsymbol{\hat{n}}$ is the
projection of the stress onto the inviscid yield surface. This equation may be
written as%
\begin{equation}
2\mu\bar{\tau}c^{2}\dot{\gamma}=f_{p}\;\Longleftrightarrow\;\dot{\gamma}%
=\frac{f_{p}}{2\mu c^{2}\bar{\tau}} \,\,\,\left( =\frac{f_{p}}{\eta
},\,\,\text{as seen below}\right)
\end{equation}
or%
\begin{equation}
f:=f_{p}-2\mu c^{2}\bar{\tau}\dot{\gamma}=0
\end{equation}
so%
\begin{equation}
f:=f_{p}-g\left(  \dot{\gamma}\right)  =0\text{ \ with }g\left(  \dot{\gamma
}\right)  =2\mu c^{2}\bar{\tau}\dot{\gamma}=c^{2}\tilde{\eta}\dot{\gamma}%
=\eta\dot{\gamma}%
\end{equation}
with $\bar{\tau}=\tilde{\eta}/2\mu=\eta/2\mu c^{2}$ (c.f. Eq. (2.7.12) in Simo
\&\ Hughes \cite{Simo1998}, and note that $\bar{\tau}\neq\hat{\tau}$ and
$\tilde{\eta}\neq\eta$). The relation between both characteristic relaxation
times is given by the term $\eta\dot{\gamma}$, as
\begin{equation}
\bar{\tau}=\frac{(2\mu+H/c^{2})}{2\mu}\hat{\tau}%
\end{equation}
which differ for the hardening case. Remarkably, with this identification,
Eqs. (\ref{Perzyna}) and (\ref{Gammadot DL}) are identical for $N=1$, so are
the models, which are also a particular case of our continuum formulation.

However, the difference between both Perzyna and Duvaut-Lions models often
refer to the ideas behind the algorithmic setting. Indeed, the immediate
implementation of Eq. (\ref{gammadot perzyna lineal}) is, integrating the
expression during the step considering the trial overstress:%
\begin{equation}
\text{(a) Trial step: }\gamma \text{ frozen and }f_p\rightarrow\,^{tr}f_p\text{; (b) Corrector step: }{\Delta\gamma}=\frac{^{tr}f_{p}}{\eta}{\Delta t}
\label{Integration Perzyna}%
\end{equation}
which gives immediately the increment in the equivalent viscoplastic strain
$\gamma$ upon knowledge of the trial inviscid plastic yield function
$^{tr}f_{p}$, which is computed in the first \textquotedblleft
predictor\textquotedblright\ sub-step keeping frozen $\gamma$ as in inviscid plasticity (recall that we argued that this partition is not consistent in the viscoplasticity case). Of course, at the end of the step
$^{t+\Delta t}f_{p}\neq0$. As long as $f_{p}>0$, the step is viscoplastic.
Equation (\ref{Integration Perzyna}) is very simple and attractive, but is
ill-conditioned for $\eta\rightarrow0$, so the inviscid case is not recovered
by the algorithm, and numerical difficulties have been reported
\cite{Peric93,Simo1998,Souza-Neto}, etc.

On the contrary, the approach given by Eq. (\ref{Gammadot DL}) considering a
relaxation of the inviscid yield function, motivates a different
implementation, taking the constant rate $\dot{\gamma}=\Delta\gamma/\Delta t$
\begin{equation}
\underset{%
\begin{array}
[c]{c}%
\eta
\end{array}
}{\underbrace{2\mu c^{2}\bar{\tau}}}\frac{\Delta\gamma}{\Delta t}=\underset{%
\begin{array}
[c]{c}%
^{t+\Delta t}f_p
\end{array}
}{\underbrace{\overbrace{~^{tr}f_{p}}^{%
\begin{array}
[c]{c}%
^{t}f_{p}+\Delta^{tr}f_{p}%
\end{array}
}\overbrace{-2\mu\Delta\gamma-H\Delta\gamma}^{%
\begin{array}
[c]{c}%
\Delta^{ct}f_p
\end{array}
}}} \label{DLform}%
\end{equation}
so factoring-out $\Delta\gamma$ ---c.f. Eq. (3.7.5) in Simo and Hughes
\cite{Simo1998} and recall the conversions explained before Eq.
(\ref{Gammadot DL})
\begin{equation}
c\Delta\gamma=\frac{^{tr}f_{p}/2\mu c}{\dfrac{\bar{\tau}}{\Delta t}+1+H/2\mu
c^{2}} \label{Deltagamma DL}%
\end{equation}
In contrast with the implementation in Eq. (\ref{Integration Perzyna}), this
form is well conditioned for $\eta\rightarrow0$ and $\bar{\tau}\rightarrow0$,
cases in which the inviscid solution is recovered. However, note that Eq.
(\ref{Deltagamma DL}) is valid only for the specific Eqs. (\ref{Gammadot DL})
or (\ref{gammadot perzyna lineal}), but not for the more general case, often
more descriptive of experimental results, of Eq. (\ref{Perzyna}) (the reason
why the implementation of Eq. (\ref{Integration Perzyna}) is preferred in most
works in the literature), and even in the linear proportional case, it does
not bring the exact solution. Interestingly, note that the actual difference
between the Perzyna and the Duvaut-Lions model is just about the integration
of the corrector contribution and the related computational algorithm, not
about any physical consideration, so they are indeed the same ``model''. Namely, Eqs. (\ref{Integration Perzyna}) and
Eq. (\ref{DLform}) just differ in the implicit consideration of the inviscid
terms in Eq. (\ref{DLform}), which are neglected in the integration in Eq.
(\ref{Integration Perzyna}). This is the reason behind its ill-conditioning
when $\eta=0$, when the inviscid terms become the only dissipative
contribution in the step.

\subsection{Consistency model}

Another model developed to solve the previous issues is the so-called
\textquotedblleft consistency\textquotedblright\ model
\cite{Wang1997,Carosio2000,Heeres2002}. In this model, using a formulation
simplified to the case at hand to facilitate comparisons, a viscoplastic yield
condition is assumed $f_{vp}\left(  \boldsymbol{\sigma},\gamma,v\right)  $,
where $\gamma$ is the consistency parameter and $v$ is another variable,
representing in many cases $\dot{\gamma}$. Consider the present case---c.f.
Eq. (\ref{fdefinition})%
\begin{equation}
f_{vp}:=c~\boldsymbol{\hat{n}}:\boldsymbol{\sigma}-~\kappa\left(
\gamma\right)  -g\left(  v\right)
\end{equation}
e.g. using the linear relations $~\kappa\left(  \gamma\right)  =$ $~\kappa
_{0}+H\gamma$ and $g\left(  v\right)  =\eta v$
\begin{equation}
f_{vp}:=c~\boldsymbol{\hat{n}}:\boldsymbol{\sigma}-~\kappa_{0}-H\gamma-\eta v
\end{equation}
The viscoplastic yield function $f_{vp}$ governs the loading/unloading
criteria as if it were a classical yield function in elastoplasticity, i.e.
follow the Kuhn-Tucker loading/unloading conditions%
\begin{equation}
f_{vp}\leq0\text{, }\dot{\gamma}>0\text{ and }\dot{\gamma}f_{vp}=0
\end{equation}
and $f_{vp}<0$ implies purely elastic behavior, regardless of the value of
$\dot{\gamma}$, see Sec. 2.2 in Heeres et al \cite{Heeres2002}. Indeed, it is
required that during loading $\dot{\gamma}=v$, and upon unloading ($f_{vp}%
<0$), then $\gamma$ remains constant. However, after unloading, during the
unloading and reloading process, the value of $v=\dot\gamma_{u}$ (subscript
standing for onset of unloading) is frozen, so after the first unloading, the
elastic domain is enlarged by $\eta v\equiv\eta\dot{\gamma}_{u}$, so
thereafter during elastic behavior $v\neq0$ whereas $\dot{\gamma}=0$, being
this the reason why the consistency model needs $v$ and $\dot{\gamma}$ (they
may take different values). Remarkably, this is the main theoretical
(practical) difference between our present proposal and the consistency model.
We emphasize that we did not make the assumption of the existence of a yield
viscoplastic surface $f_{vp}$. Our function $f=0$ is just a power balance
which guarantees the fulfillment of the first principle of thermodynamics.
Then, our loading/unloading condition is simply given by the value of
$\dot{\gamma}$. As long as $\dot{\gamma}>0$, viscoplastic flow takes place;
the absence of it (elastic loading) requires $\dot{\gamma}=0$, and we do not
need the additional variable $v$. As a consequence, our model behaves as the
Perzyna model, whereas the consistency model reloads to the previous unloading
stress-strain point, as noted in Heeres et al \cite{Heeres2002}, see therein
Figures 1 and 3. Also noteworthy, the Perzyna and the consistency models give
the same results if no unloading takes place, see Figs. 4 and 6 in Heeres et
al \cite{Heeres2002}. We mention that the inclusion of the possibility of
using $\eta\rightarrow0$ with more general viscoplastic constitutive equations
of the type Eq. (\ref{Perzyna}) comes with the cost of a more complex
algorithm, e.g. Sec. 4.2 in Heeres et al \cite{Heeres2002}.

\subsection{Models without yield function. Nonlinear viscoelasticity}

Many models, as the Norton-Odqvist law, do not employ a yield function (i.e. a
yield stress). This implies that the viscoplastic strain is given directly in
terms of the stress, e.g. Norton's law is%
\begin{equation}
\dot{\gamma}=\left(  \frac{c\left\Vert \boldsymbol{\sigma}^{d}\right\Vert
}{\breve{\eta}}\right)  ^{N}\label{Eq_Norton}%
\end{equation}
with $\left\Vert \boldsymbol{\sigma}^{d}\right\Vert \equiv\boldsymbol{\hat{n}%
}:\boldsymbol{\sigma}$ and%
\begin{equation}
-~^{ct}\boldsymbol{\dot{\varepsilon}}_{e}=c\dot{\gamma}\boldsymbol{\hat{n}%
}\label{ee}%
\end{equation}
Norton's law can be written, taking $\kappa\left(  \gamma\right)  =0$, as%
\begin{equation}
f\equiv f_{p}-g\left(  \dot{\gamma}\right)  \equiv c\mathbf{\hat{n}%
}:\boldsymbol{\sigma}-\breve{\eta}\dot{\gamma}^{1/N}=0
\end{equation}
Then, if we just take $f_{p}=c\boldsymbol{\hat{n}}:\boldsymbol{\sigma}$
and $g\left(  \dot{\gamma}\right)  =\breve{\eta}\dot{\gamma}^{1/N}$, and
$g^{\prime}\left(  \dot{\gamma}\right)  =\tfrac{1}{N}\breve\eta\dot{\gamma}^{1/N-1}$,
our formulation and integration algorithm are unchanged and
well-conditioned, being this just a particular case. Indeed, the absence of
yield stress is the case of viscoelasticity, so the present formulation
recovers naturally the viscoelasticity formulation as a particular case; see
e.g. \cite{LatMonViscoCAS,LatMonViscoCM}. Note that all equations are valid
just setting $\kappa\left(  \gamma\right)  =0$, e.g. Eqs. (\ref{inviscidrate}%
), (\ref{gammadot}) and (\ref{dotgamma}), and that the evolution equation in
Refs. \cite{LatMonViscoCAS,LatMonViscoCM} is, for the linear isotropic case
considered therein%
\begin{equation}
-~^{ct}\boldsymbol{\dot{\varepsilon}}_{e}=c^{2}\eta^{-1}\boldsymbol{\sigma
}^{d}%
\end{equation}
which is just a reformatting of Eqs. (\ref{Eq_Norton}) and (\ref{ee}).

\subsection{Models with kinematic hardening}

The friction element in the rehological model has only isotropic hardening.
However, the formulation is essentially valid for kinematic hardening,
including the nonlinear kinematic hardening case (e.g. Ohno-Wang model). To
this end, it only suffices to include a spring in parallel to the Bingham
model, and include in the formulation the corresponding stored energy (note
that kinematic hardening has energetic nature). This setting also holds in the
case of large strains employing the Kr\"{o}ner-Lee multiplicative
decomposition. For more details on this type of formulation see Refs.
\cite{LatMonPlas,SanzLatMon,Sanzetal,NguyenSanzMontans,ZhangMontans}

\section{Uniaxial numerical comparisons with classical models for linear
viscoplasticiy}

In this section we compare the results against other formulations (models and
algorithms). We consider in this case the homogeneous, proportional linear
case under loading and reverse loading to highlight similarities and
differences, as often performed in the literature. A single integration point
is subjected to an infinitesimal shear load with a constant shear strain rate.
For this comparison, the proposed model and other three well-known models
(Perzyna, Duvault-Lions and consistency model) are implemented. Different
values of shear strain rate and different values of the time increment are
also applied in order to analyze the influence on the viscoplastic response
and on the accuracy. The constitutive material parameters are given in Table
\ref{table.1}. For the other models, the proper equivalence, presented in the
previous sections, are employed.

\begin{table}[h]
\caption{Constitutive parameters for the viscoplastic model}
\centering%
\begin{tabular}
[c]{ccccc}\hline
$E$ [kPa] & $\nu$ & $\eta$ [kPa s] & $\kappa_{0}$ [kPa] & $H$ [kPa]\\\hline
$2.0E07$ & $0.2$ & $2.0E03$ & $2.0E03$ & $5.0E06$\\\hline
\end{tabular}
\label{table.1}%
\end{table}

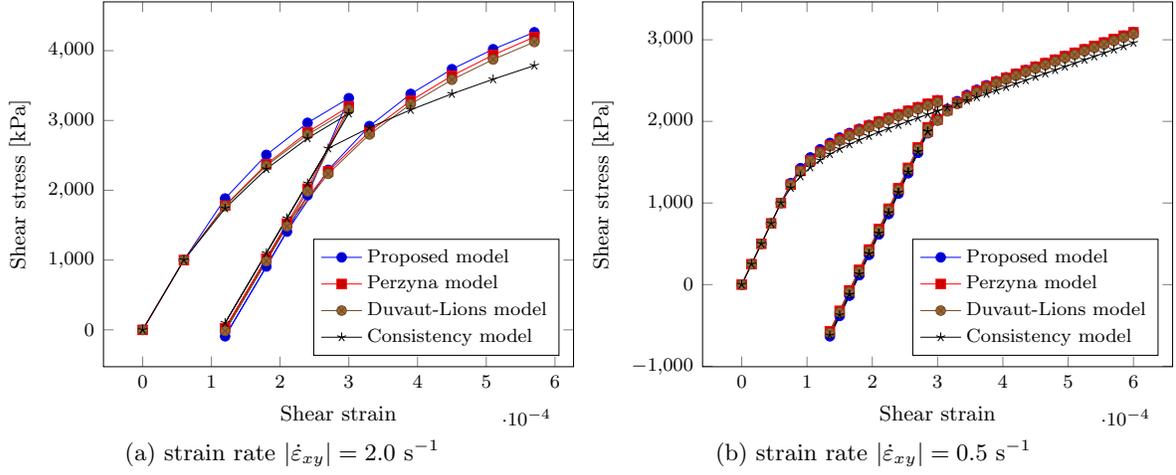
\begin{figure}[pth]
\centering
\subfloat[strain rate $|\dot{\varepsilon}_{xy}| =2.0$ s$^{-1}$ ]{
\begin{tikzpicture}[scale = 0.9]
\pgfplotsset{every axis legend/.append style={at={(0.5,1.03)},anchor=south east},}
\begin{axis}[legend style ={cells={anchor=west},legend pos=south east}, ylabel={Shear stress [kPa]},
xlabel={Shear strain}]
\addplot  table [x expr=\thisrowno{1}, y expr=\thisrowno{2}]{results_pure_shear_dt_3.0e-5_reps_2.0_eta_1.0_loading-unloading.txt};
\addplot  table [x expr=\thisrowno{1}, y expr=\thisrowno{3}]{results_pure_shear_dt_3.0e-5_reps_2.0_eta_1.0_loading-unloading.txt};
\addplot  table [x expr=\thisrowno{1}, y expr=\thisrowno{4}]{results_pure_shear_dt_3.0e-5_reps_2.0_eta_1.0_loading-unloading.txt};
\addplot  table [x expr=\thisrowno{1}, y expr=\thisrowno{5}]{results_pure_shear_dt_3.0e-5_reps_2.0_eta_1.0_loading-unloading.txt};
\legend{Proposed model, Perzyna model, Duvaut-Lions model, Consistency model}
\end{axis}
\end{tikzpicture}}
%-----
\subfloat[strain rate $|\dot{\varepsilon}_{xy}| = 0.5$ s$^{-1}$]{
\begin{tikzpicture}[scale = 0.9]
\pgfplotsset{every axis legend/.append style={at={(0.5,1.03)},anchor=south east},}
\begin{axis}[legend style ={cells={anchor=west},legend pos=south east}, ylabel={Shear stress [kPa]},
xlabel={Shear strain}]
\addplot  table [x expr=\thisrowno{1}, y expr=\thisrowno{2}]{results_pure_shear_dt_3.0e-5_reps_0.5_eta_1.0_loading-unloading.txt};
\addplot  table [x expr=\thisrowno{1}, y expr=\thisrowno{3}]{results_pure_shear_dt_3.0e-5_reps_0.5_eta_1.0_loading-unloading.txt};
\addplot  table [x expr=\thisrowno{1}, y expr=\thisrowno{4}]{results_pure_shear_dt_3.0e-5_reps_0.5_eta_1.0_loading-unloading.txt};
\addplot  table [x expr=\thisrowno{1}, y expr=\thisrowno{5}]{results_pure_shear_dt_3.0e-5_reps_0.5_eta_1.0_loading-unloading.txt};
\legend{Proposed model, Perzyna model, Duvaut-Lions model, Consistency model}
\end{axis}
\end{tikzpicture}}\caption{Loading-unloading-reloading with different models
for $\Delta t=3.0\times10^{-5}$ s}%
\label{fig.comparison1}%
\end{figure}

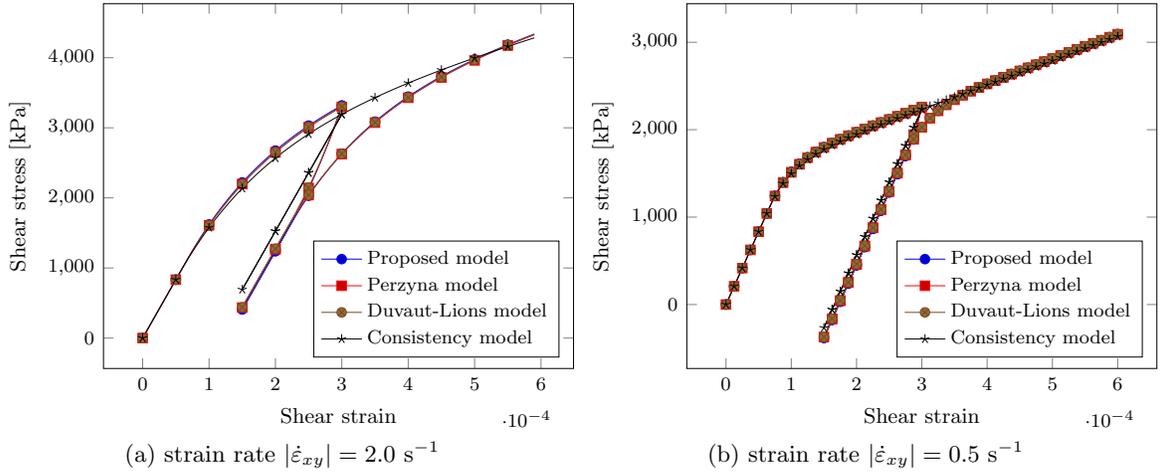
\begin{figure}[pth]
\centering
\subfloat[strain rate $|\dot{\varepsilon}_{xy}| = 2.0$ s$^{-1}$]{
\begin{tikzpicture}[scale = 0.9]
\pgfplotsset{every axis legend/.append style={at={(0.5,1.03)},anchor=south east},}
\begin{axis}[legend style ={cells={anchor=west},legend pos=south east}, ylabel={Shear stress [kPa]},
xlabel={Shear strain}, mark repeat = 5]
\addplot  table [x expr=\thisrowno{1}, y expr=\thisrowno{2}]{results_pure_shear_dt_5.0e-6_reps_2.0_eta_1.0_loading-unloading.txt};
\addplot  table [x expr=\thisrowno{1}, y expr=\thisrowno{3}]{results_pure_shear_dt_5.0e-6_reps_2.0_eta_1.0_loading-unloading.txt};
\addplot  table [x expr=\thisrowno{1}, y expr=\thisrowno{4}]{results_pure_shear_dt_5.0e-6_reps_2.0_eta_1.0_loading-unloading.txt};
\addplot  table [x expr=\thisrowno{1}, y expr=\thisrowno{5}]{results_pure_shear_dt_5.0e-6_reps_2.0_eta_1.0_loading-unloading.txt};
\legend{Proposed model, Perzyna model, Duvaut-Lions model, Consistency model}
\end{axis}
\end{tikzpicture}}
%-----
\subfloat[strain rate $|\dot{\varepsilon}_{xy}| = 0.5$ s$^{-1}$]{
\begin{tikzpicture}[scale = 0.9]
\pgfplotsset{every axis legend/.append style={at={(0.5,1.03)},anchor=south east},}
\begin{axis}[legend style ={cells={anchor=west},legend pos=south east}, ylabel={Shear stress [kPa]},
xlabel={Shear strain}, mark repeat = 5]
\addplot  table [x expr=\thisrowno{1}, y expr=\thisrowno{2}]{results_pure_shear_dt_5.0e-6_reps_0.5_eta_1.0_loading-unloading.txt};
\addplot  table [x expr=\thisrowno{1}, y expr=\thisrowno{3}]{results_pure_shear_dt_5.0e-6_reps_0.5_eta_1.0_loading-unloading.txt};
\addplot  table [x expr=\thisrowno{1}, y expr=\thisrowno{4}]{results_pure_shear_dt_5.0e-6_reps_0.5_eta_1.0_loading-unloading.txt};
\addplot  table [x expr=\thisrowno{1}, y expr=\thisrowno{5}]{results_pure_shear_dt_5.0e-6_reps_0.5_eta_1.0_loading-unloading.txt};
\legend{Proposed model, Perzyna model, Duvaut-Lions model, Consistency model}
\end{axis}
\end{tikzpicture}}
%----
\caption{Loading-unloading-reloading with different models for $\Delta
t=5.0\times10^{-6}$ s}%
\label{fig.comparison2}%
\end{figure}

Figures \ref{fig.comparison1} and \ref{fig.comparison2} show the comparison
including stress reversals for different shear strain rate and different time
increments. The stress reversals consist of an initial loading phase (up to a
shear strain of $3.0\times10^{-4}$), then an unloading phase is applied until
shear strain of $1.5\times10^{-4}$, and finally a reloading phase is followed
to a shear strain of $6.0\times10^{-4}$. It can be observed that the
significant difference in the viscoplastic behavior of different models starts
from the moment of crossing the limit of the elastic domain at $\kappa
_{0}/\sqrt{3}=1,154.7%
%TCIMACRO{\unit{kPa}}%
%BeginExpansion
\operatorname{kPa}%
%EndExpansion
$. A noticeable difference can be seen during the unloading phase. Our
proposed consistency viscoplasticity model uses the viscoplastic multiplier
rate ($\dot{\gamma}$) to check whether dissipation occurs. Therefore,
dissipation, and hence viscoplastic deformation, is produced as long as
$\dot{\gamma}>0$. Our model unloads elastically when $\dot{\gamma}$ just
vanishes. This behavior is similar to that of both the Perzyna model and the
Duvaut-Lions model due to the effect known as \textquotedblleft
overstress\textquotedblright. On the contrary, the consistency model
\cite{Wang1997,Carosio2000,Heeres2002} always unloads elastically because the
dynamic loading surface is treated as a yield surface, enclosing an elastic
domain. This different unloading behavior also leads to a noticeable
difference in the subsequent reloading phase.

Also noticeable is that the Perzyna and Duvaut-Lions models related
integration algorithms give results close to those of our model for small
strain rates, when an accurate integration of the rate $\dot{\gamma}$ (and
hence of the dynamic contribution) is not so relevant (e.g. the case for
$|\dot{\varepsilon}_{xy}|=0.5/s$ ); in the loading phase a similar result is
also observed with the consistency model in
\cite{Wang1997,Carosio2000,Heeres2002}. However, as strain rate increases and
the dynamic contribution becomes more relevant, the difference between models
is more noticeable. Indeed, unlike other models, since our model gives the
exact solution for this linear case, the strain rate \emph{does not} affect
the accuracy of our predictions, and in turn, this accuracy is not affected by
the time increment of the step.

Another relevant difference is observed for the consistency model, which is
apparent specially in Fig. \ref{fig.comparison1}a. During the reloading phase,
the trial value of $f_{vp\text{ }}$ governs the instant when the step becomes
fully viscoplastic (even if there is an initial fraction which would be
elastic). Then, because of this numerical inaccuracy, it regains viscoplastic
behavior before reaching the previous unloading stress point. Of course for
small steps, this effect becomes negligible, see Figs. \ref{fig.comparison1}b
and \ref{fig.comparison2}.

%----------------------

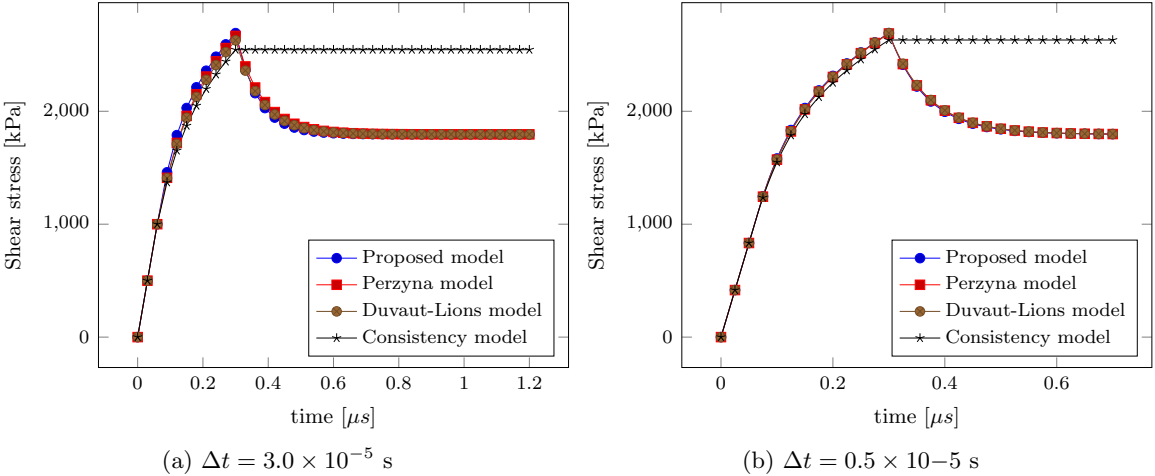
\begin{figure}[pth]
\centering
\subfloat[$\Delta t=3.0\times10^{-5}$ s]{
\begin{tikzpicture}[scale = 0.9]
\pgfplotsset{every axis legend/.append style={at={(0.5,1.03)},anchor=south east},}
\begin{axis}[legend style ={cells={anchor=west},legend pos=south east}, ylabel={Shear stress [kPa]},
xlabel={time [$\mu s$]}]
\addplot  table [x expr=1000*\thisrowno{0}, y expr=\thisrowno{2}]{results_pure_shear_dt_3.0e-5_reps_1.0_eta_1.0_relaxation.txt};
\addplot  table [x expr=1000*\thisrowno{0}, y expr=\thisrowno{3}]{results_pure_shear_dt_3.0e-5_reps_1.0_eta_1.0_relaxation.txt};
\addplot  table [x expr=1000*\thisrowno{0}, y expr=\thisrowno{4}]{results_pure_shear_dt_3.0e-5_reps_1.0_eta_1.0_relaxation.txt};
\addplot  table [x expr=1000*\thisrowno{0}, y expr=\thisrowno{5}]{results_pure_shear_dt_3.0e-5_reps_1.0_eta_1.0_relaxation.txt};
\legend{Proposed model, Perzyna model, Duvaut-Lions model, Consistency model}
\end{axis}
\end{tikzpicture}}
%-----
\subfloat[$\Delta t=0.5\times 10{-5}$ s]{
\begin{tikzpicture}[scale = 0.9]
\pgfplotsset{every axis legend/.append style={at={(0.5,1.03)},anchor=south east},}
\begin{axis}[legend style ={cells={anchor=west},legend pos=south east}, ylabel={Shear stress [kPa]},
xlabel={time [$\mu s$]}, mark repeat=5]
\addplot  table [x expr=1000*\thisrowno{0}, y expr=\thisrowno{2}]{results_pure_shear_dt_5.0e-6_reps_1.0_eta_1.0_relaxation.txt};
\addplot  table [x expr=1000*\thisrowno{0}, y expr=\thisrowno{3}]{results_pure_shear_dt_5.0e-6_reps_1.0_eta_1.0_relaxation.txt};
\addplot  table [x expr=1000*\thisrowno{0}, y expr=\thisrowno{4}]{results_pure_shear_dt_5.0e-6_reps_1.0_eta_1.0_relaxation.txt};
\addplot  table [x expr=1000*\thisrowno{0}, y expr=\thisrowno{5}]{results_pure_shear_dt_5.0e-6_reps_1.0_eta_1.0_relaxation.txt};
\legend{Proposed model, Perzyna model, Duvaut-Lions model, Consistency model}
\end{axis}
\end{tikzpicture}}\caption{Relaxation: shear stress decay at constant strain
with different models}%
\label{fig.comparison3}%
\end{figure}

The relaxation behavior of all models is also analysed. To this end, the
shear strain is increased employing a constant shear strain rate
$\dot{\varepsilon}_{xy}=1.0/s$ to achieve a maximum shear strain of
$3.0\times10^{-4}$, and thereafter is left constant. Figure
\ref{fig.comparison3} shows the results of this simulation for different time
increments. Again, for small step sizes, the Perzyna model shows the same
response as our proposed model.

\begin{figure}[pth]
\centering
\begin{tikzpicture}
\pgfplotsset{every axis legend/.append style={at={(0.5,1.03)},anchor=south east},}
\begin{axis}[legend style ={cells={anchor=west},legend pos=south east}, ylabel={Shear stress [kPa]},
xlabel={Shear strain}, mark repeat = 5]
\addplot  table [x expr=\thisrowno{0}, y expr=\thisrowno{2}]{results_pure_shear_dt_5.0e-6_reps_1.0_eta_0.0_loading.txt};
\addplot  table [x expr=\thisrowno{0}, y expr=\thisrowno{3}]{results_pure_shear_dt_5.0e-6_reps_1.0_eta_0.0_loading.txt};
\addplot  table [x expr=\thisrowno{0}, y expr=\thisrowno{4}]{results_pure_shear_dt_5.0e-6_reps_1.0_eta_0.0_loading.txt};
\addplot  table [x expr=\thisrowno{0}, y expr=\thisrowno{5}]{results_pure_shear_dt_5.0e-6_reps_1.0_eta_0.0_loading.txt};
\legend{Proposed model, Perzyna model, Duvaut-Lions model, Consistency model}
\end{axis}
\end{tikzpicture}
\caption{Shear stress - strain curves for viscosity $\eta=0$}%
\label{fig.eta0}%
\end{figure}
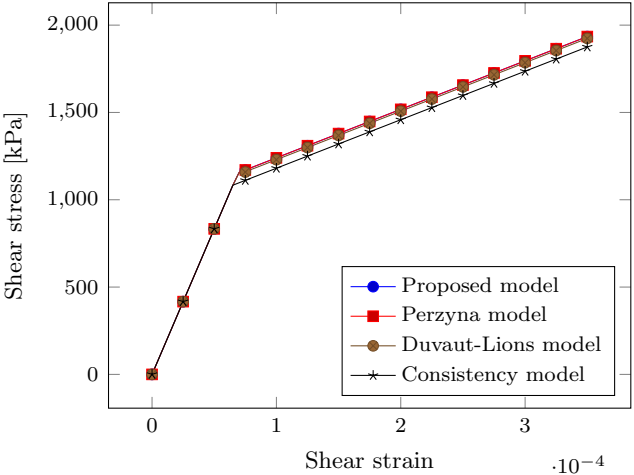

Furthermore, in order to check the accurate performance of our proposed
consistency viscoplasticity model we have performed a numerical testing
for the case of $\eta=0.0$, i.e. totally elasto-plastic model. Figure
\ref{fig.eta0} represents a comparison of results obtained by different
models. To avoid the ill-conditioning of the Perzyna model for $\eta
\rightarrow0$, a small $\eta$ and quadruple (real*64 type) precision has been
employed. It can be seen that all models give the same solution for zero
viscosity, except for the consistency model. Again, this difference is due to
the use of the trial $f_{vp}$ value to detect a viscoplastic step and consider
it fully viscoplastic (note that the initial error is just maintained during
the rest of the simulation, and vanishes when using small $\Delta t$).

\section{General discrete formulation: A simple backward-Euler integration
algorithm for non-constant material parameters}

In this section we introduce a general formulation for the nonlinear
viscoelasticity case, which obviously recovers the aforediscussed linear
formulation as a particular case.

\subsection{Local Newton algorithm for the fully viscous step}

We have shown in the previous examples that the linear model is capable of
recovering the exact solution in the proportional case. In most practical
loading cases at a stress point in a finite element simulation, the loading in
a step is almost proportional (meaning that the change in the direction of
$\mathbf{\hat{n}}$ is small). Then, it seems reasonable to develop an
integration algorithm that recovers that exact solution for the linear
proportional case. To this end, in contrast with typical viscoplasticity
algorithms, two independent variables are considered at each step, namely
$^{t}\gamma$ and $^{t}\dot{\gamma}$. Two conditions are enforced for the
integration algorithm. The first one is the preservation of energy for
$\dot{\gamma}\neq0$; i.e. Eq. (\ref{conditions})$_{1}$ if $^{t+\Delta t}%
\dot{\gamma}\neq0$%
\begin{align}
^{t+\Delta t}r_{f}\left(  ^{t+\Delta t}\gamma,~^{t+\Delta t}\dot{\gamma
}\right)   &  =c~^{t+\Delta t}\boldsymbol{\sigma}:~^{t+\Delta t}%
\hat{\boldsymbol{n}}-\kappa\left(  ^{t+\Delta t}\gamma\right)  -g\left(
^{t+\Delta t}\dot{\gamma}\right)  \\
&  =\underset{%
\begin{array}
[c]{c}%
^{tr}f
\end{array}
}{\underbrace{c~^{tr}\boldsymbol{\sigma}:~^{tr}\hat{\boldsymbol{n}}%
-~^{t}\kappa-~^{t}g}}\underset{%
\begin{array}
[c]{c}%
\Delta^{ct}f
\end{array}
}{\underbrace{-c^{2}\Delta\gamma~\mathbb{C}_{e}:~^{tr}\hat{\boldsymbol{n}%
}-\Delta\kappa-\Delta g}}\rightarrow0
\end{align}
with $\Delta\left(  \cdot\right)  =~^{t+\Delta t}\left(  \cdot\right)
-~^{t}\left(  \cdot\right)  $ and note that as in perfect plasticity,
$~^{t+\Delta t}\hat{\boldsymbol{n}}=~^{tr}\hat{\boldsymbol{n}}=~^{tr}%
\boldsymbol{\sigma}^{d}\boldsymbol{/}\left\Vert ~^{tr}\boldsymbol{\sigma}%
^{d}\right\Vert $. The second one gives the relation for the conservation
during the step, which relates $\gamma$ and $\dot{\gamma}$ through the
solution of the corresponding differential Equation (\ref{ODE2_v1}), starting
at $\gamma_{0}=~^{t}\gamma$ and ending at $\gamma=~^{t+\Delta t}\gamma$, with
a constant $\boldsymbol{\dot{\varepsilon}}=\Delta\boldsymbol{\varepsilon
}/\Delta t$ during the step---c.f. Eq. (\ref{gammadot2})%
\begin{equation}
^{t+\Delta t}r_{\dot{\gamma}}\left(  ^{t+\Delta t}\gamma,~^{t+\Delta t}%
\dot{\gamma}\right)  :=~^{t+\Delta t}\dot{\gamma}-~^{t+\Delta t}\dot{\gamma
}_{\infty}-\left[  ^{t}\dot{\gamma}-~^{t+\Delta t}\dot{\gamma}_{\infty
}\right]  \exp\left(  -\frac{\Delta t}{^{t+\Delta t}\hat{\tau}}\right)
\rightarrow0
\end{equation}
In the previous expressions%
\begin{equation}
\left\{
\begin{array}
[c]{l}%
^{t+\Delta t}\hat{\tau}\left(  ^{t+\Delta t}\dot{\gamma},^{t+\Delta t}%
\gamma\right)  :=g^{\prime}(^{t+\Delta t}\dot{\gamma})~/~^{t+\Delta
t}h(^{t+\Delta t}\gamma)\\
\;\\
^{t+\Delta t}\dot{\gamma}_{\infty}\left(  \Delta\boldsymbol{\varepsilon
},^{t+\Delta t}\gamma\right)  :=\left[  c~^{t+\Delta t}\boldsymbol{\hat{n}%
}:\mathbb{C}_{e}:\Delta\boldsymbol{\varepsilon}/\Delta t\right]  ~/~^{t+\Delta
t}h(^{t+\Delta t}\gamma)\\
\;\\
^{t+\Delta t}h(^{t+\Delta t}\gamma):=c^{2}~^{t+\Delta t}\boldsymbol{\hat{n}%
}:\mathbb{C}_{e}:~^{t+\Delta t}\boldsymbol{\hat{n}}+\kappa^{\prime}(^{t+\Delta
t}\gamma)\\
\;\\
~^{t+\Delta t}\boldsymbol{\sigma}\left(  ~^{t+\Delta t}\boldsymbol{\varepsilon
}_{e}\left(  \Delta\boldsymbol{\varepsilon},^{t+\Delta t}\gamma\right)
\right)  =\mathbb{C}_{e}:~^{t+\Delta t}\boldsymbol{\varepsilon}_{e}\left(
\Delta\boldsymbol{\varepsilon},^{t+\Delta t}\gamma,^{t+\Delta t}\dot{\gamma
}\right)  \\
\;\\
~^{t+\Delta t}\boldsymbol{\hat{n}}\left(  \Delta\boldsymbol{\varepsilon
}\right)  =~^{t+\Delta t}\boldsymbol{\sigma}^{d}/~\left\Vert ~^{t+\Delta
t}\boldsymbol{\sigma}^{d}\right\Vert =~^{tr}\boldsymbol{\sigma}^{d}\left(
\Delta\boldsymbol{\varepsilon}\right)  /~\left\Vert ~^{tr}\boldsymbol{\sigma
}^{d}\left(  \Delta\boldsymbol{\varepsilon}\right)  \right\Vert
\end{array}
\right.
\end{equation}
where we declared explicitly the dependencies for further reference and note
that $~^{t+\Delta t}\boldsymbol{\hat{n}}:\mathbb{C}_{e}:~^{t+\Delta
t}\boldsymbol{\hat{n}}=2\mu$ (constant). The two residues can be written in
vector form as%

\begin{equation}
\mathsf{R}(\mathsf{E})=\left\{
\begin{matrix}
r_{\dot{\gamma}}\\
r_{f}%
\end{matrix}
\right\}  \rightarrow\bm {0}\quad\text{with}\quad\mathsf{E}=\left\{
\begin{matrix}
{}^{t+\Delta t}\dot{\gamma}\\
^{t+\Delta t}\gamma
\end{matrix}
\right\}
\end{equation}
The residual vector equation is solved using the Newton-Raphson method, where
the solution is updated at iteration $(j+1)$ from the known values at
iteration $(j)$ by%

\begin{equation}
\mathsf{E}^{(j+1)}=\mathsf{E}^{(j)}-[\nabla\mathsf{R}^{(j)}]^{-1}%
\mathsf{R}^{(j)}\label{local NR}%
\end{equation}
until
\begin{equation}
||\mathsf{R}^{(j+1)}||\leq{tol}%
\end{equation}
where $(\bullet)^{\left[  j\right]  }$ indicates quantities for iteration
$\left[  j\right]  $ at time step $t+\Delta t$. For the first iteration, we
take $\mathsf{E}_{i}^{^{\left[  0\right]  }}=[~^{t}\dot{\gamma},~^{t}%
\gamma]^{T}$ and the trial value $~^{tr}\boldsymbol{\sigma}=\mathbb{C}%
_{e}:\left(  ~^{t}\boldsymbol{\varepsilon}_{e}+\Delta\boldsymbol{\varepsilon
}\right)  $. The Jacobian of the residual vector respect to the variables is%
\begin{equation}
\nabla\mathsf{R}^{\left[  j\right]  }=%
\begin{bmatrix}
\dfrac{\partial^{t+\Delta t}r_{\dot{\gamma}}^{\left[  j\right]  }}%
{\partial^{t+\Delta t}\dot{\gamma}} & \dfrac{\partial^{t+\Delta t}%
r_{\dot{\gamma}}^{\left[  j\right]  }}{\partial^{t+\Delta t}\gamma}\\
\dfrac{\partial^{t+\Delta t}r_{f}^{\left[  j\right]  }}{\partial^{t+\Delta
t}\dot{\gamma}} & \dfrac{\partial^{t+\Delta t}r_{f}^{\left[  j\right]  }%
}{\partial^{t+\Delta t}\gamma}%
\end{bmatrix}
\end{equation}
The first derivative is%
\begin{equation}
\dfrac{\partial^{t+\Delta t}r_{\dot{\gamma}}}{\partial^{t+\Delta t}\dot
{\gamma}}=1-\frac{^{t}\dot{\gamma}-~^{t+\Delta t}\dot{\gamma}_{\infty}%
}{^{t+\Delta t}\hat{\tau}^{2}}\frac{d^{t+\Delta t}\hat{\tau}}{\partial
^{t+\Delta t}\dot{\gamma}}\exp\left(  -\frac{\Delta t}{^{t+\Delta t}\hat{\tau
}}\right)
\end{equation}
with%
\begin{equation}
\frac{\partial^{t+\Delta t}\hat{\tau}}{\partial^{t+\Delta t}\dot{\gamma}%
}:=\frac{g^{\prime\prime}(^{t+\Delta t}\dot{\gamma})}{~^{t+\Delta t}h}%
-\frac{g^{\prime}(^{t+\Delta t}\dot{\gamma})}{~^{t+\Delta t}h^{2}}%
\frac{\partial^{t+\Delta t}h}{\partial^{t+\Delta t}\dot{\gamma}}%
\end{equation}
so%
\begin{equation}
\nabla\mathsf{R}_{11}\equiv\;\dfrac{\partial^{t+\Delta t}r_{\dot{\gamma}}%
}{\partial^{t+\Delta t}\dot{\gamma}}=1-\frac{^{t}\dot{\gamma}-~^{t+\Delta
t}\dot{\gamma}_{\infty}}{^{t+\Delta t}\hat{\tau}^{2}}\frac{g^{\prime\prime
}(^{t+\Delta t}\dot{\gamma})}{~^{t+\Delta t}h}\exp\left(  -\frac{\Delta
t}{^{t+\Delta t}\hat{\tau}}\right)
\end{equation}
where we used $\partial~^{t+\Delta t}\boldsymbol{\hat{n}/}\partial^{t+\Delta
t}\dot{\gamma}=\mathbf{0}$, and $\partial^{t+\Delta t}h/\partial^{t+\Delta
t}\dot{\gamma}=0$ and $\partial^{t+\Delta t}\dot{\gamma}_{\infty}%
/\partial^{t+\Delta t}\dot{\gamma}=0$ (because they only depend on $^{t+\Delta
t}\gamma$) The second derivative is%
\begin{equation}
\dfrac{\partial^{t+\Delta t}r_{\dot{\gamma}}}{\partial^{t+\Delta t}\gamma
}=-\frac{\partial^{t+\Delta t}\dot{\gamma}_{\infty}}{\partial^{t+\Delta
t}\gamma}\left[  1-\exp\left(  -\frac{\Delta t}{^{t+\Delta t}\hat{\tau}%
}\right)  \right]  -\frac{^{t}\dot{\gamma}-~^{t+\Delta t}\dot{\gamma}_{\infty
}}{^{t+\Delta t}\hat{\tau}^{2}}\frac{d^{t+\Delta t}\hat{\tau}}{\partial
^{t+\Delta t}\gamma}\exp\left(  -\frac{\Delta t}{^{t+\Delta t}\hat{\tau}%
}\right)
\end{equation}
with $\partial^{t+\Delta t}h/\partial^{t+\Delta t}\gamma=\kappa^{\prime\prime
}(^{t+\Delta t}\gamma)$ and%
\begin{equation}
\frac{\partial~^{t+\Delta t}\dot{\gamma}_{\infty}}{\partial^{t+\Delta t}%
\gamma}=-\frac{c~^{t+\Delta t}\boldsymbol{\hat{n}}:\mathbb{C}_{e}%
:\Delta\mathbf{\varepsilon}/\Delta t}{~^{t+\Delta t}h^{2}}\kappa^{\prime
\prime}(^{t+\Delta t}\gamma)
\end{equation}
and%
\[
\frac{d^{t+\Delta t}\hat{\tau}}{\partial^{t+\Delta t}\gamma}=-\frac{g^{\prime
}(^{t+\Delta t}\dot{\gamma})}{~^{t+\Delta t}h^{2}}~\kappa^{\prime\prime
}(^{t+\Delta t}\gamma)
\]
so%
\begin{align}
\left.  \nabla\mathsf{R}_{12}\equiv\;\dfrac{\partial^{t+\Delta t}%
r_{\dot{\gamma}}}{\partial^{t+\Delta t}\gamma}\right.   &  =c~^{t+\Delta
t}\boldsymbol{\hat{n}}:\mathbb{C}_{e}:\frac{\Delta\mathbf{\varepsilon}}{\Delta
t}\frac{\kappa^{\prime\prime}(^{t+\Delta t}\gamma)}{~^{t+\Delta t}h^{2}%
}\left[  1-\exp\left(  -\frac{\Delta t}{^{t+\Delta t}\hat{\tau}}\right)
\right]  \nonumber\\
&  +\frac{^{t}\dot{\gamma}-~^{t+\Delta t}\dot{\gamma}_{\infty}}{^{t+\Delta
t}\hat{\tau}^{2}}g^{\prime}(^{t+\Delta t}\dot{\gamma})~\frac{\kappa
^{\prime\prime}(^{t+\Delta t}\gamma)}{~^{t+\Delta t}h^{2}}\exp\left(
-\frac{\Delta t}{^{t+\Delta t}\hat{\tau}}\right)
\end{align}
and note that since $\Delta\gamma$ does not change the return direction (as
previously anticipated) we get (and have used this result in the previous
equations) $\partial~^{t+\Delta t}\boldsymbol{\hat{n}/}\partial^{t+\Delta
t}\gamma=\mathbf{0}$.

For the third derivative, note that $^{tr}f$ does not depend on $^{t+\Delta
t}\dot{\gamma}$ nor on $^{t+\Delta t}\gamma$, so using the previous results,
is%
\begin{equation}
\left.  \nabla\mathsf{R}_{21}\equiv\;\dfrac{\partial^{t+\Delta t}r_{f}%
}{\partial^{t+\Delta t}\gamma}\right.  =-c^{2}~^{t+\Delta t}\hat
{\boldsymbol{n}}:\mathbb{C}_{e}:~^{t+\Delta t}\hat{\boldsymbol{n}}%
-\kappa^{\prime}\left(  ^{t+\Delta t}\gamma\right)
\end{equation}
where we used%
\begin{equation}
\frac{\partial^{t+\Delta t}\boldsymbol{\sigma}}{\partial^{t+\Delta t}\gamma
}=\mathbb{C}_{e}:\frac{\partial^{t+\Delta t}\boldsymbol{\varepsilon}_{e}%
}{\partial^{t+\Delta t}\gamma}=-c\mathbb{C}_{e}:~^{t+\Delta t}\hat
{\boldsymbol{n}}%
\end{equation}
Finally, the fourth derivative, taking again into account the previous
results, is%
\begin{equation}
\nabla\mathsf{R}_{22}\equiv\;\dfrac{\partial^{t+\Delta t}r_{f}}{\partial
^{t+\Delta t}\dot{\gamma}}=-g^{\prime}\left(  ^{t+\Delta t}\dot{\gamma
}\right)
\end{equation}

Obviously, for the linear proportional case, we must recover the exact
solution explained in the previous sections. In this case%
\begin{equation}
\left\{
\begin{array}
[c]{l}%
\nabla\mathsf{R}_{11}\equiv\;\dfrac{\partial^{t+\Delta t}r_{\dot{\gamma}}%
}{\partial^{t+\Delta t}\dot{\gamma}}=1-g^{\prime\prime}(^{t+\Delta t}%
\dot{\gamma})\left(  ...\right)  =1\\
\nabla\mathsf{R}_{12}\equiv\;\dfrac{\partial^{t+\Delta t}r_{\dot{\gamma}}%
}{\partial^{t+\Delta t}\gamma}=\kappa^{\prime\prime}(^{t+\Delta t}%
\gamma)\left(  ...\right)  =0\\
\nabla\mathsf{R}_{21}\equiv\;\dfrac{\partial^{t+\Delta t}r_{f}}{\partial
^{t+\Delta t}\gamma}=-2\mu c^{2}-H\\
\nabla\mathsf{R}_{22}\equiv\;\dfrac{\partial^{t+\Delta t}r_{f}}{\partial
^{t+\Delta t}\dot{\gamma}}=-\eta
\end{array}
\right.
\end{equation}
so inverting the matrix and solving for just an iteration (note that
$\,^{t+\Delta t}\dot{\gamma}_{\infty}$ is explicitly known at this point)%
\begin{equation}
\left[
\begin{array}
[c]{c}%
^{t+\Delta t}\dot{\gamma}\\
^{t+\Delta t}\gamma
\end{array}
\right]  =\left[
\begin{array}
[c]{c}%
^{t}\dot{\gamma}\\
^{t}\gamma
\end{array}
\right]  +\left[
\begin{array}
[c]{cc}%
1 & 0\\
\tau & \left(  2\mu c^{2}+H\right)  ^{-1}%
\end{array}
\right]  \left[
\begin{array}
[c]{c}%
\left(  ^{t}\dot{\gamma}-~^{t+\Delta t}\dot{\gamma}_{\infty}\right)  \left[
1-\exp\left(  -\dfrac{\Delta t}{\hat{\tau}}\right)  \right]  \\
^{tr}f
\end{array}
\right]
\end{equation}
we recover the solution for the linear case, in a well-conditioned manner
(regardless of the value of $\eta$), as expected, see Eqs. (\ref{gammadotn+1})
and (\ref{dgamman+1}).

\subsection{Tangent for global equilibrium iterations}

In deriving the tangent, when changing the strain increment $\Delta
\boldsymbol{\varepsilon}$ we must guarantee that the two conditions
$r_{\dot{\gamma}}=0$ and $r_{f}=0$ still hold. This means that, upon local
convergence%
\begin{equation}
\left\{
\begin{array}
[c]{c}%
d^{t+\Delta t}r_{f}\left(  \Delta\boldsymbol{\varepsilon}\mathbf{,}^{t+\Delta
t}\gamma\left(  \Delta\boldsymbol{\varepsilon}\right)  ,~^{t+\Delta t}%
\dot{\gamma}\left(  \Delta\boldsymbol{\varepsilon}\right)  \right)
~/~d\Delta\boldsymbol{\varepsilon}=0\\
d^{t+\Delta t}r_{\dot{\gamma}}\left(  \Delta\boldsymbol{\varepsilon}%
\mathbf{,}^{t+\Delta t}\gamma\left(  \Delta\boldsymbol{\varepsilon}\right)
,~^{t+\Delta t}\dot{\gamma}\left(  \Delta\boldsymbol{\varepsilon}\right)
\right)  ~/~d\Delta\boldsymbol{\varepsilon}=0
\end{array}
\right.
\end{equation}
The rates are%
\begin{align}
\frac{d^{t+\Delta t}r_{f}}{~d\Delta\boldsymbol{\varepsilon}} &  =\frac
{\partial^{t+\Delta t}r_{f}}{~\partial\Delta\boldsymbol{\varepsilon}}%
+\frac{\partial^{t+\Delta t}r_{f}}{~\partial^{t+\Delta t}\gamma}\frac
{\partial^{t+\Delta t}\gamma}{~\partial\Delta\boldsymbol{\varepsilon}}%
+\frac{\partial^{t+\Delta t}r_{f}}{~\partial^{t+\Delta t}\dot{\gamma}}%
\frac{\partial^{t+\Delta t}\dot{\gamma}}{~\partial\Delta
\boldsymbol{\varepsilon}}\nonumber\\
&  =\underset{%
\begin{array}
[c]{c}%
d^{tr}f/d\Delta\boldsymbol{\varepsilon}%
\end{array}
}{\underbrace{c~^{t+\Delta t}\hat{\boldsymbol{n}}:\mathbb{C}_{e}%
+~^{tr}\boldsymbol{\sigma}^{d}:\mathbb{P}_{n}:\mathbb{C}_{e}}}+\underset{%
\begin{array}
[c]{c}%
d\Delta^{ct}f/d\Delta\boldsymbol{\varepsilon}%
\end{array}
}{\underbrace{\nabla\mathsf{R}_{22}\frac{\partial^{t+\Delta t}\gamma
}{~\partial\Delta\boldsymbol{\varepsilon}}+\nabla\mathsf{R}_{21}\frac
{\partial^{t+\Delta t}\dot{\gamma}}{~\partial\Delta\boldsymbol{\varepsilon}}}%
}=\mathbf{0}%
\end{align}%
\begin{align}
\frac{d^{t+\Delta t}r_{\dot{\gamma}}}{~d\Delta\boldsymbol{\varepsilon}} &
=\frac{\partial^{t+\Delta t}r_{\dot{\gamma}}}{\partial\Delta
\boldsymbol{\varepsilon}}+\frac{\partial^{t+\Delta t}r_{\dot{\gamma}}%
}{~\partial^{t+\Delta t}\gamma}\frac{\partial^{t+\Delta t}\gamma}%
{~\partial\Delta\boldsymbol{\varepsilon}}+\frac{\partial^{t+\Delta t}%
r_{\dot{\gamma}}}{~\partial^{t+\Delta t}\dot{\gamma}}\frac{\partial^{t+\Delta
t}\dot{\gamma}}{~\partial\Delta\boldsymbol{\varepsilon}}\nonumber\\
&  =\frac{\partial^{t+\Delta t}r_{\dot{\gamma}}}{\partial\Delta
\mathbf{\varepsilon}}+\nabla\mathsf{R}_{12}\frac{\partial^{t+\Delta t}\gamma
}{~\partial\Delta\boldsymbol{\varepsilon}}+\nabla\mathsf{R}_{11}\frac
{\partial^{t+\Delta t}\dot{\gamma}}{~\partial\Delta\boldsymbol{\varepsilon}%
}\nonumber\\
&  =-~\frac{\partial^{t+\Delta t}\dot{\gamma}_{\infty}}{~\partial
\Delta\boldsymbol{\varepsilon}}\left[  1-\exp\left(  -\frac{\Delta
t}{^{t+\Delta t}\hat{\tau}}\right)  \right]  +\nabla\mathsf{R}_{11}%
\frac{\partial^{t+\Delta t}\gamma}{~\partial\Delta\boldsymbol{\varepsilon}%
}+\nabla\mathsf{R}_{12}\frac{\partial^{t+\Delta t}\dot{\gamma}}{~\partial
\Delta\boldsymbol{\varepsilon}}=\mathbf{0}%
\end{align}
with%
\begin{align}
\frac{\partial^{t+\Delta t}\dot{\gamma}_{\infty}\left(  \Delta
\boldsymbol{\varepsilon},^{t+\Delta t}\gamma\right)  }{~\partial
\Delta\boldsymbol{\varepsilon}} &  =\frac{c}{\Delta t~^{t+\Delta
t}h(^{t+\Delta t}\gamma)}\left[  ~^{tr}\boldsymbol{\hat{n}}:\mathbb{C}%
_{e}+\frac{\Delta\boldsymbol{\varepsilon}^{d}}{\left\Vert ~^{tr}%
\boldsymbol{n}\right\Vert }:\mathbb{C}_{e}:\mathbb{P}_{n}:\mathbb{C}%
_{e}\right]  \nonumber\\
&  =\frac{c}{\Delta t~^{t+\Delta t}h(^{t+\Delta t}\gamma)}\left[  2\mu
~^{tr}\boldsymbol{\hat{n}}+\frac{\left(  2\mu\right)  ^{2}}{\left\Vert
~^{tr}\boldsymbol{n}\right\Vert }\mathbb{P}_{n}:\Delta\boldsymbol{\varepsilon
}^{d}\right]
\end{align}
Both conditions give immediately the quantities $\partial^{t+\Delta t}%
\gamma/~\partial\Delta\mathbf{\varepsilon}$ and $\partial^{t+\Delta t}%
\gamma/~\partial\Delta\mathbf{\varepsilon}$ by solving the system of equations%
\begin{equation}
\left[
\begin{array}
[c]{cc}%
\nabla\mathsf{R}_{11} & \nabla\mathsf{R}_{12}\\
\nabla\mathsf{R}_{21} & \nabla\mathsf{R}_{22}%
\end{array}
\right]  \left[
\begin{array}
[c]{c}%
\left(  \partial^{t+\Delta t}\dot{\gamma}/\partial\Delta
\boldsymbol{\varepsilon}\right)  ^{T}\\
\left(  \partial^{t+\Delta t}\gamma/\partial\Delta\boldsymbol{\varepsilon
}\right)  ^{T}%
\end{array}
\right]  =-\left[
\begin{array}
[c]{c}%
-\left(  ~\partial^{t+\Delta t}\dot{\gamma}_{\infty}/~\partial\Delta
\boldsymbol{\varepsilon}\right)  ^{T}\left[  1-\exp\left(  -\dfrac{\Delta
t}{^{t+\Delta t}\hat{\tau}}\right)  \right]  \\
\left(  c~^{t+\Delta t}\hat{\boldsymbol{n}}:\mathbb{C}_{e}+~^{tr}%
\boldsymbol{\sigma}^{d}:\mathbb{P}_{n}:\mathbb{C}_{e}\right)  ^{T}%
\end{array}
\right]
\end{equation}
The stress tensor is given by%
\begin{equation}
~^{t+\Delta t}\boldsymbol{\sigma}\left(  ~^{t+\Delta t}\boldsymbol{\varepsilon
}_{e}\left(  \Delta\boldsymbol{\varepsilon},^{t+\Delta t}\gamma\right)
\right)  =\mathbb{C}_{e}:~^{t+\Delta t}\boldsymbol{\varepsilon}_{e}%
=\mathbb{C}_{e}:\left[  ~^{t}\boldsymbol{\varepsilon}_{e}+\Delta
\boldsymbol{\varepsilon}-\left(  ^{t+\Delta t}\gamma-~^{t}\gamma\right)
~^{tr}\boldsymbol{\hat{n}}\right]
\end{equation}
so%
\begin{equation}
\mathbb{C}:=\frac{d~^{t+\Delta t}\boldsymbol{\sigma}\left(  ~^{t+\Delta
t}\boldsymbol{\varepsilon}_{e}\left(  \Delta\boldsymbol{\varepsilon
},^{t+\Delta t}\gamma\left(  \Delta\boldsymbol{\varepsilon}\right)  \right)
\right)  }{d\Delta\boldsymbol{\varepsilon}}=\mathbb{C}_{e}:\left[
~\mathbb{I}^{S}-~^{tr}\boldsymbol{\hat{n}}\mathbf{\otimes}\frac{\partial
^{t+\Delta t}\gamma}{\partial\Delta\boldsymbol{\varepsilon}}-\frac
{\Delta\gamma}{\left\Vert ~^{tr}\boldsymbol{n}\right\Vert }\mathbb{P}%
_{n}:\mathbb{C}_{e}\right]
\end{equation}
of which all quantities are known. We note that in contrast to that reported
in \cite{Heeres2002} (see Sec. 4.2 therein), our algorithmic tangent is
symmetric, the same way as those of inviscid plasticity and viscoelasticity.

\subsection{Inviscid to viscous case}

In this case it is also possible to compute the $\Delta t^{c}$ such that the
inviscid plastic yield surface is crossed. The procedure is similar to that
developed for the linear case. However, assuming that we are dealing with a
nonlinear case in which the solution will be approximate, a simpler acceptable
procedure may be to just consider the step as fully viscoplastic (as e.g. in
\cite{Heeres2002}), in which initially $^{t}\dot{\gamma}=0$ and in which
finally $^{t+\Delta t}f=0$. If the steps are small, the error induced in this
step will also be small. Note that in this case, the consistent tangent for
equilibrium iterations is the same as in the previous case.

\subsection{Viscous to inviscid case}

This case is detected by a result of a trial fully viscoplastic step in which
$^{t+\Delta t}\dot{\gamma}<0$ and/or $^{t+\Delta t}f_{p}\leq0$ (note that
because of the approximations in the nonlinear case, it is possible that both
conditions are not met simultaneously). Then, for example, when the condition
$^{t+\Delta t}\dot{\gamma}<0$ is detected, the step may be considered elastic,
by simply setting $^{t+\Delta t}\dot{\gamma}=0$ and approximating
$\Delta\gamma$ by (in small steps it can also be taken $\Delta\gamma\simeq0$)%
\begin{equation}
\Delta\gamma\simeq\frac{^{t}f_{p}}{~^{t}h(^{t}\gamma+\hat{\tau}^{t}\dot
{\gamma})}%
\end{equation}

In this case, the consistent tangent for equilibrium iterations is the elastic
one $\mathbb{C}_{e}$.

\section{Numerical examples}

\begin{figure}[pth]
\centering
\includegraphics[width=0.5\textwidth]{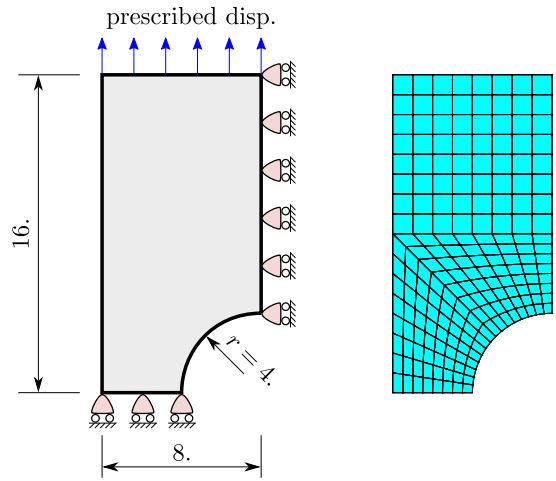}\caption{Strip with
circular hole: geometry and FE model}%
\label{fig.plate_hole}%
\end{figure}

The purpose of this section is to show the numerical performance of the
proposed algorithm in a typical finite element simulation using a nonlinear
viscoplasticity model. A Perzyna-type nonlinear model for $g\left(
\dot{\gamma}\right)  $ is employed, see Eq. (\ref{eqfPerzyna}), so for the
viscous contribution, a viscosity $\bar{\eta}$ and the rate sensitivity
parameter $N$ are used (apart from the adimensionalyzing parameter
$\bar{\kappa}=\kappa_{0}$). For the inviscid part $\kappa\left(
\gamma\right)  $, a Voce-type relation is employed. The parameters of the
model are given in Table \ref{table.3}. The numerical example consists in the
extension of a strip of thickness $1$ mm with a central circular hole. It is
presented to assess accuracy and robustness of the proposed viscoplasticity model and of the adopted numerical scheme. The strip is subjected
to a loading simulated via imposed displacement in the vertical direction up
to $u/L=0.32%
%TCIMACRO{\unit{mm}}%
%BeginExpansion
\operatorname{mm}%
%EndExpansion
/16%
%TCIMACRO{\unit{mm}}%
%BeginExpansion
\operatorname{mm}%
%EndExpansion
$, where $u$ is the prescribed displacement and $L$ is the length of the
specimen, see Fig. \ref{fig.plate_hole}. The considered loading rates $\dot
{u}/L$ are $2.0/s$, $1.0/s$ and $0.5/s$. The analysis corresponds to only one
quarter part of the plate, taking into account its symmetries. Figure
\ref{fig.plate_hole} shows the geometry and the finite element discretization.
High order mixed $u/p$ fully integrated ($3\times3\times3$ Gauss integration)
Q2/P1 - 27/4 brick finite elements are used for this analysis.

\begin{table}[h]
\caption{Material parameters for the plate hole simulation. Voce's law is
$\kappa\left(  \gamma\right)  =\kappa_{0}+H\gamma+\left(  \kappa_{\infty
}-\kappa_{0}\right)  \exp\left(  -\delta\gamma\right)  $. Perzyna's model
employed is $\dot{\gamma}=\left\langle f_{p}/\kappa_{0}\right\rangle ^{N}%
/\bar{\eta}$, so $g\left(  \dot{\gamma}\right)  =\kappa_{0}\bar{\eta}%
^{1/N}\dot{\gamma}^{1/N}$.}
\centering%
\begin{tabular}
[c]{ll}\hline
Young modulus & $E=206.9$ GPa\\
Poisson coef. & $\nu=0.29$\\
Reference yield stress & $\kappa_{0}=450.0$ MPa\\
Limit stress parameter & $\kappa_{\infty}=550.0$ MPa\\
Hardening modulus & $\overline{H}=200.0$ MPa\\
Voce exponential parameter & $\delta=10$\\
Viscosity parameter & $\bar{\eta}=1$ s\\\hline
\end{tabular}
\label{table.3}%
\end{table}

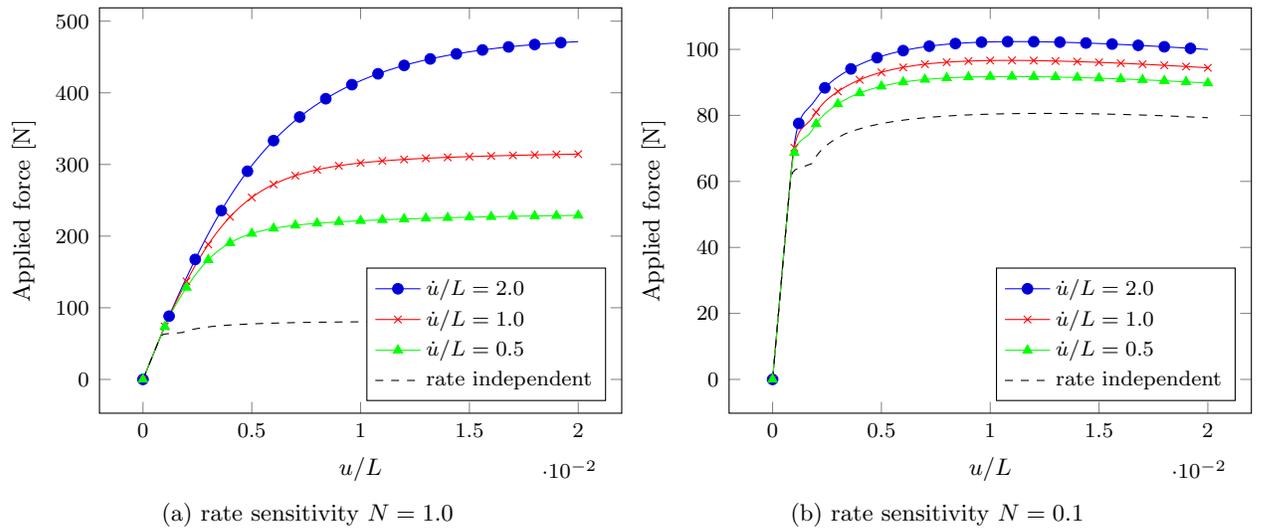
\begin{figure}[pth]
\centering
\subfloat[rate sensitivity $N=1.0$]{
\begin{tikzpicture}
\pgfplotsset{every axis legend/.append style={at={(0.5,1.03)},anchor=south east},}
\begin{axis}[legend style ={cells={anchor=west},legend pos=south east}, ylabel={Applied force [N]},
xlabel={$u/L$},mark repeat = 30]
\addplot  table [x expr=0.0625*\thisrowno{0}, y expr=-\thisrowno{1}]{fuerza_reaccion_rate2.0.dat};
\addplot[color=red, mark=x, mark repeat = 50]  table [x expr=0.0625*\thisrowno{0}, y expr=-\thisrowno{1}]{fuerza_reaccion_rate1.0.dat};
\addplot[color=green, mark=triangle*, mark repeat = 100]   table [x expr=0.0625*\thisrowno{0}, y expr=-\thisrowno{1}]{fuerza_reaccion_rate0.5.dat};
\addplot[color=black, dashed, mark repeat = 50]   table [x expr=0.0625*\thisrowno{0}, y expr=-\thisrowno{1}]{fuerza_reaccion_rate_independent.dat};
\legend{$\dot{u}/L = 2.0$, $\dot{u}/L = 1.0$, $\dot{u}/L = 0.5$, rate independent}
\end{axis}
\end{tikzpicture}}
%---
\subfloat[rate sensitivity $N=0.1$]{
\begin{tikzpicture}
\pgfplotsset{every axis legend/.append style={at={(0.5,1.03)},anchor=south east},}
\begin{axis}[legend style ={cells={anchor=west},legend pos=south east}, ylabel={Applied force [N]},
xlabel={$u/L$},mark repeat = 30]
\addplot  table [x expr=0.0625*\thisrowno{0}, y expr=-\thisrowno{1}]{fuerza_reaccion_rate2.0_sensi_0.1.dat};
\addplot[color=red, mark=x, mark repeat = 50]  table [x expr=0.0625*\thisrowno{0}, y expr=-\thisrowno{1}]{fuerza_reaccion_rate1.0_sensi_0.1.dat};
\addplot[color=green, mark=triangle*, mark repeat = 100]   table [x expr=0.0625*\thisrowno{0}, y expr=-\thisrowno{1}]{fuerza_reaccion_rate0.5_sensi_0.1.dat};
\addplot[color=black, dashed, mark repeat = 50]   table [x expr=0.0625*\thisrowno{0}, y expr=-\thisrowno{1}]{fuerza_reaccion_rate_independent.dat};
\legend{$\dot{u}/L = 2.0$, $\dot{u}/L = 1.0$, $\dot{u}/L = 0.5$, rate independent}
\end{axis}
\end{tikzpicture}}
%---
\caption{Force vesus displacement curves for different loading rates}%
\label{fig.force_displ_1}%
\end{figure}

The numerical solutions are obtained with our in-house finite element code
Dulcinea. Figure \ref{fig.force_displ_1} gives a comparison of the
force-displacement curve between results obtained by changing the loading rate
for two values of rate sensitivity $\bar{N}=1.0$ (a linear viscoplastic case)
and $N=0.1$ (nonlinear viscoplastic case). The major influence of the loading
rate is observed for the high rate-sensitive material as expected. For the low
loading rate $\dot{u}/L=0.5$ and for low rate sensitivity $N=0.1$, the
obtained solution tends to the rate-independent solution, as expected.

\begin{figure}[pth]
\centering
\includegraphics[width=\textwidth]{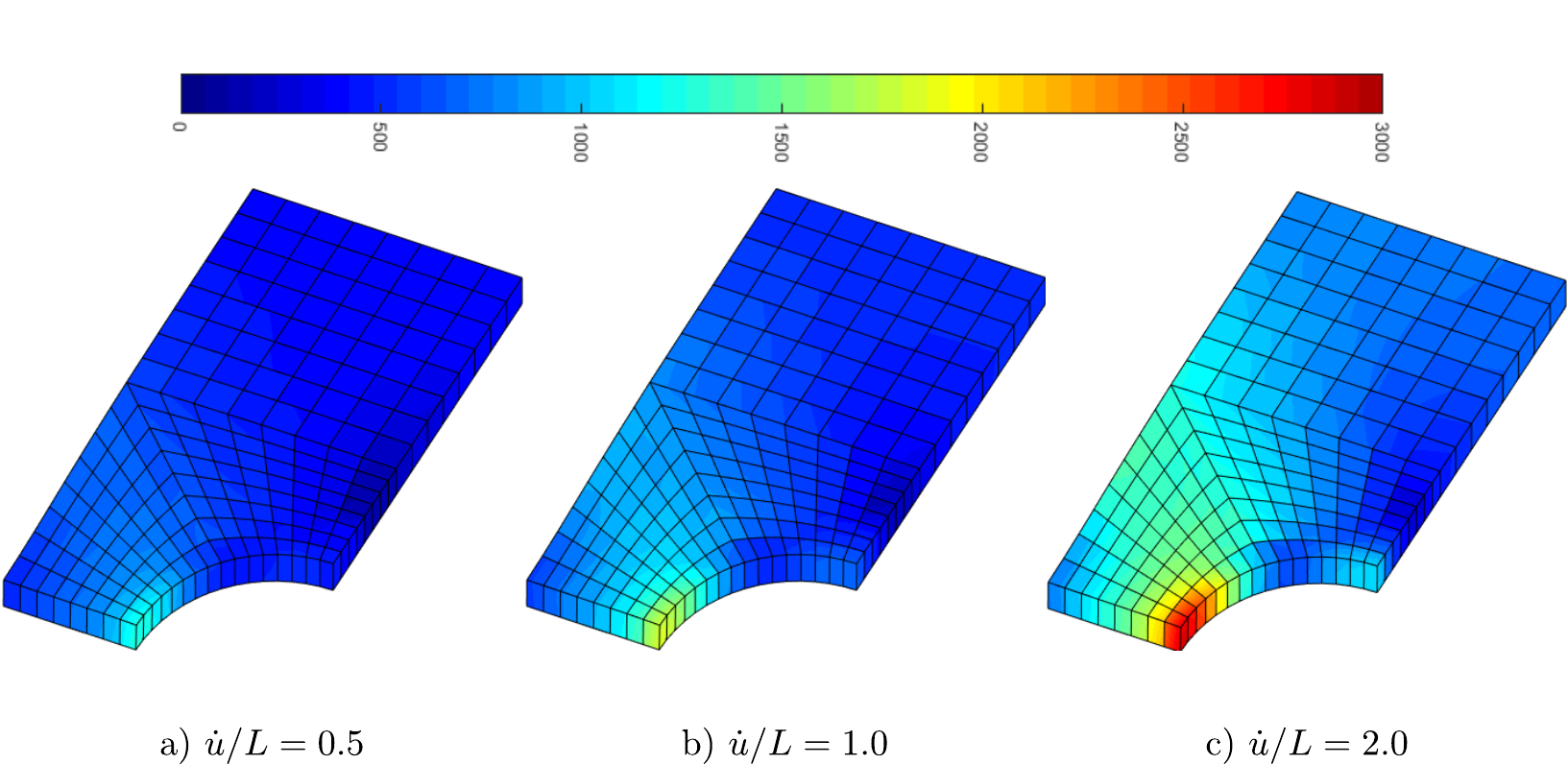}\caption{Stretching
of the strip with circular hole: von Mises stress at final prescribed
displacement for different loading rate.}%
\label{fig.vonMises_2}%
\end{figure}

The von Mises stress contour at the final prescribed displacement is shown in
Figure \ref{fig.vonMises_2} for different loading rates. A higher von Mises
stress level is observed for high loading rates as expected and it is
in concordance with the force-displacement curve presented in Figure
\ref{fig.force_displ_1}. Both local and global convergence rates are
asymptotically quadratic, and typical values are given in Table
\ref{tab.convergence_rate1} for different steps.

\begin{table}[h]
\caption{Extension of strip with circular hole. Convergence rates for global
Newton-Raphson iterations (rnom= residual in force; enorm= residual in
energy; R($\bullet$)= error in local equations)}%
\label{tab.convergence_rate1}
\centering
\begin{tabular}
[c]{ccccc}%
\multicolumn{5}{c}{Global convergence}\\\hline
\multirow{2}{*}{Iteration} & \multicolumn{2}{c}{Step 50} &
\multicolumn{2}{c}{Step 200}\\\cline{2-5}
& rnorm & enorm & rnorm & enorm\\
\midrule 1 & $1.188E+02$ & $1.571E-01$ & $1.191E+02$ & $1.631E-01$\\
2 & $5.133E-03$ & $1.627E-10$ & $1.118E-03$ & $2.049E-11$\\
3 & $8.058E-07$ & $8.522E-18$ & $6.957E-06$ & $6.891E-16$\\\hline
\multicolumn{5}{c}{Local convergence}\\\hline
Iteration & \multicolumn{2}{c}{R(1)} & \multicolumn{2}{c}{R(2)}\\
1 & \multicolumn{2}{c}{$1.323E+01$} & \multicolumn{2}{c}{$3.924E+01$}\\
2 & \multicolumn{2}{c}{$9.691E-03$} & \multicolumn{2}{c}{$3.017E-02$}\\
3 & \multicolumn{2}{c}{$4.425E-10$} & \multicolumn{2}{c}{$4.133E-09$}\\
4 & \multicolumn{2}{c}{$1.013E-14$} & \multicolumn{2}{c}{$1.319E-13$}\\
\bottomrule &  &  &  &
\end{tabular}
\end{table}

\section{Conclusions}

In this work we present a novel treatment of viscoplasticity, both
from a theoretical side and a computational one. One of our purposes has been
to integrate exactly the linear proportional case in a manner such that the
viscous behavior is constructed from the inviscid one in rate form, the latter
recovered automatically for vanishing viscosities. However, we pursued a
formulation also valid for more general nonlinear viscoplasticity cases which,
furthermore, recovers the visco\emph{elastic} formulation for vanishing
yield surfaces. The formulation unifies naturally the plasticity, viscoelasticity and
viscoplasticity models and algorithms.

Essential to the developments has been the derivation of the evolution
equations from thermodynamics, considering separately the conservation of
power from the conservation of energy, the former yielding a constitutive
equation for the equivalent viscoplastic strain rate, and the second one
giving an extra equation for the computation of the equivalent viscoplastic
strain. In the linear proportional case, the solution is exact for a given
step. However, this setting also allows for a simple incorporation of the
general nonlinear viscoplasticity models.

We have presented and analyzed the model and integration procedure using a
small strains framework based on elastic corrector rates. As we have shown in
previous works in anisotropic elastoplasticity and viscoelasticity, this
framework can be easily extended to large strains employing classical
multiplicative decompositions and logarithmic strains, still resulting in the
same additive structure, and reducing large strains to kinematic pre- and post-processors.

\section*{Acknowledgments}

Partial financial support for this work has been given by Agencia Estatal de
Investigaci\'{o}n of Spain under grant PGC2018-097257-B-C32.

\bibliography{references}

\end{document}